\documentclass[twocolumn,twocolappendix]{aastex631}

\usepackage{multirow}

\newcommand{\hst}{\textit{HST}}
\newcommand{\spitzer}{\textit{Spitzer}}
\newcommand{\jwst}{\textit{JWST}}
\newcommand{\herschel}{\textit{Herschel}}

\newcommand{\hb}{\hbox{H$\beta$}}
\newcommand{\ha}{\hbox{H$\alpha$}}
\newcommand{\galfit}{\hbox{\textsc{galfit}}}
\newcommand{\cigale}{\hbox{\textsc{cigale}}}

\begin{document}

\title{CEERS: Spatially Resolved UV and mid-IR Star Formation in Galaxies at $0.2 < z < 2.5$:\\ The Picture from the \textit{Hubble} and \textit{James Webb} Space Telescopes}

%\correspondingauthor{August Muench}
%\email{greg.schwarz@aas.org, gus.muench@aas.org}

\correspondingauthor{Lu Shen}
\email{lushen@ustc.edu.cn}

\author[0000-0001-9495-7759]{Lu Shen}
\affiliation{Department of Physics and Astronomy, Texas A\&M University, College
Station, TX, 77843-4242 USA}
\affiliation{George P.\ and Cynthia Woods Mitchell Institute for
 Fundamental Physics and Astronomy, Texas A\&M University, College
 Station, TX, 77843-4242 USA}
 
\author[0000-0001-7503-8482]{Casey Papovich}
\affiliation{Department of Physics and Astronomy, Texas A\&M University, College
Station, TX, 77843-4242 USA}
\affiliation{George P.\ and Cynthia Woods Mitchell Institute for
 Fundamental Physics and Astronomy, Texas A\&M University, College
 Station, TX, 77843-4242 USA}

\author[0000-0001-8835-7722]{Guang Yang}
\affiliation{Kapteyn Astronomical Institute, University of Groningen, P.O. Box 800, 9700 AV Groningen, The Netherlands}
\affiliation{SRON Netherlands Institute for Space Research, Postbus 800, 9700 AV Groningen, The Netherlands}

\author[0000-0002-7547-3385]{Jasleen Matharu}
\affiliation{Cosmic Dawn Center, Niels Bohr Institute, University of Copenhagen, R\aa dmandsgade 62, 2200 Copenhagen, Denmark\\}

\author[0000-0002-9373-3865]{Xin Wang}
\affiliation{School of Astronomy and Space Science, University of Chinese Academy of Sciences (UCAS), Beijing 100049, China}
\affiliation{National Astronomical Observatories, Chinese Academy of Sciences, Beijing 100101, China}
\affiliation{Institute for Frontiers in Astronomy and Astrophysics, Beijing Normal University,  Beijing 102206, China}

\author[0000-0002-6777-6490]{Benjamin Magnelli}
\affiliation{Universit{\'e} Paris-Saclay, Universit{\'e} Paris Cit{\'e}, CEA, CNRS, AIM, 91191, Gif-sur-Yvette, France}

\author[0000-0002-7631-647X]{David Elbaz}
\affiliation{Universit{\'e} Paris-Saclay, Universit{\'e} Paris Cit{\'e}, CEA, CNRS, AIM, 91191, Gif-sur-Yvette, France}

\author[0000-0002-1590-0568]{Shardha Jogee}
\affiliation{Department of Astronomy, The University of Texas at Austin, Austin, TX, USA}

% UVCANDELS + CEERS + etc in alphabetical order

\author[0000-0002-8630-6435]{Anahita Alavi}
\affiliation{IPAC, Mail Code 314-6, California Institute of Technology, 1200 E. California Blvd., Pasadena, CA, 91125, USA}

\author[0000-0002-7959-8783]{Pablo Arrabal Haro}
\affiliation{NSF's National Optical-Infrared Astronomy Research Laboratory, 950 N. Cherry Ave., Tucson, AZ 85719, USA}

\author[0000-0001-8534-7502]{Bren E. Backhaus}
\affiliation{Department of Physics, 196 Auditorium Road, Unit 3046, University of Connecticut, Storrs, CT 06269}

\author[0000-0002-9921-9218]{Micaela B. Bagley}
\affiliation{Department of Astronomy, The University of Texas at Austin, Austin, TX, USA}

\author[0000-0002-5564-9873]{Eric F.\ Bell}
\affiliation{Department of Astronomy, University of Michigan, 1085 S. University Ave, Ann Arbor, MI 48109-1107, USA}

\author[0000-0003-0492-4924]{Laura Bisigello}
\affiliation{Dipartimento di Fisica e Astronomia "G.Galilei", Universit\'a di Padova, Via Marzolo 8, I-35131 Padova, Italy}
\affiliation{INAF--Osservatorio Astronomico di Padova, Vicolo dell'Osservatorio 5, I-35122, Padova, Italy}

\author[0000-0003-2536-1614]{Antonello Calabr{\`o}} 
\affiliation{INAF - Osservatorio Astronomico di Roma, via di Frascati 33, 00078 Monte Porzio Catone, Italy}

\author[0000-0003-1371-6019]{M. C. Cooper}
\affiliation{Department of Physics \& Astronomy, University of California, Irvine, 4129 Reines Hall, Irvine, CA 92697, USA}

\author[0000-0001-6820-0015]{Luca Costantin}
\affiliation{Centro de Astrobiolog\'ia (CSIC-INTA), Ctra de Ajalvir km 4, Torrej\'on de Ardoz, 28850, Madrid, Spain}

\author[0000-0002-3331-9590]{Emanuele Daddi}
\affiliation{Universit{\'e} Paris-Saclay, Universit{\'e} Paris Cit{\'e}, CEA, CNRS, AIM, 91191, Gif-sur-Yvette, France}

\author[0000-0001-5414-5131]{Mark Dickinson}
\affiliation{NSF's National Optical-Infrared Astronomy Research Laboratory, 950 N. Cherry Ave., Tucson, AZ 85719, USA}

\author[0000-0001-8519-1130]{Steven L. Finkelstein}
\affiliation{Department of Astronomy, The University of Texas at Austin, Austin, TX, USA}

\author[0000-0001-7201-5066]{Seiji Fujimoto}
\affiliation{Cosmic Dawn Center (DAWN), Jagtvej 128, DK2200 Copenhagen N, Denmark}
\affiliation{Niels Bohr Institute, University of Copenhagen, Lyngbyvej 2, DK2100 Copenhagen \O, Denmark}

\author[0000-0002-7831-8751]{Mauro Giavalisco}
\affiliation{University of Massachusetts Amherst, 710 North Pleasant Street, Amherst, MA 01003-9305, USA}

\author[0000-0001-9440-8872]{Norman A. Grogin}
\affiliation{Space Telescope Science Institute, 3700 San Martin Dr., Baltimore, MD 21218, USA}

\author[0000-0002-4162-6523]{Yuchen Guo}
\affiliation{Department of Astronomy, The University of Texas at Austin, Austin, TX, USA}

\author[0000-0002-4884-6756]{Benne W. Holwerda}
\affil{Physics \& Astronomy Department, University of Louisville, 40292 KY, Louisville, USA}

\author[0000-0001-9187-3605]{Jeyhan S. Kartaltepe}
\affiliation{Laboratory for Multiwavelength Astrophysics, School of Physics and Astronomy, Rochester Institute of Technology, 84 Lomb Memorial Drive, Rochester, NY 14623, USA}

\author[0000-0002-6610-2048]{Anton M. Koekemoer}
\affiliation{Space Telescope Science Institute, 3700 San Martin Dr., Baltimore, MD 21218, USA}

\author[0000-0002-8816-5146]{Peter Kurczynski}
\affiliation{Goddard Space Flight Center, 8800 Greenbelt Rd., Greenbelt MD 20771}

\author[0000-0003-1581-7825]{Ray A. Lucas}
\affiliation{Space Telescope Science Institute, 3700 San Martin Drive, Baltimore, MD 21218, USA}

\author[0000-0003-4528-5639]{Pablo G. P\'erez-Gonz\'alez}
\affiliation{Centro de Astrobiolog\'{\i}a (CAB), CSIC-INTA, Ctra. de Ajalvir km 4, Torrej\'on de Ardoz, E-28850, Madrid, Spain}

\author[0000-0003-3382-5941]{Nor Pirzkal}
\affiliation{ESA/AURA Space Telescope Science Institute}

\author[0000-0002-0604-654X]{Laura Prichard}
\affiliation{Space Telescope Science Institute, 3700 San Martin Dr., Baltimore, MD 21218, USA}

\author[0000-0002-9946-4731]{Marc Rafelski}
\affiliation{Space Telescope Science Institute, 3700 San Martin Drive, Baltimore, MD 21218, USA}
\affiliation{Department of Physics and Astronomy, Johns Hopkins University, Baltimore, MD 21218, USA}

\author[0000-0001-5749-5452]{Kaila Ronayne}
\affiliation{Department of Physics and Astronomy, Texas A\&M University, College
Station, TX, 77843-4242 USA}
\affiliation{George P.\ and Cynthia Woods Mitchell Institute for
 Fundamental Physics and Astronomy, Texas A\&M University, College
 Station, TX, 77843-4242 USA}

\author[0000-0002-6386-7299]{Raymond C. Simons}
\affiliation{Department of Physics, 196 Auditorium Road, Unit 3046, University of Connecticut, Storrs, CT 06269, USA}

\author[0000-0003-3759-8707]{Ben Sunnquist}
\affiliation{Space Telescope Science Institute, 3700 San Martin Dr., Baltimore, MD 21218, USA}

\author[0000-0002-7064-5424]{Harry I. Teplitz}
\affiliation{IPAC, Mail Code 314-6, California Institute of Technology, 1200 E. California Blvd., Pasadena CA, 91125, USA}

\author[0000-0002-1410-0470]{Jonathan R. Trump}
\affiliation{Department of Physics, 196 Auditorium Road, Unit 3046, University of Connecticut, Storrs, CT 06269, USA}

\author[0000-0001-6065-7483]{Benjamin J. Weiner}
\affiliation{MMT/Steward Observatory, University of Arizona, 933 N. Cherry Ave., Tucson, AZ 85721, USA}

\author[0000-0001-8156-6281]{Rogier A. Windhorst} %%% Rogier.Windhorst@gmail.com
\affiliation{School of Earth and Space Exploration, Arizona State University,
Tempe, AZ 85287-1404, USA}

\author[0000-0003-3466-035X]{{L. Y. Aaron} {Yung}}
\affiliation{Astrophysics Science Division, NASA Goddard Space Flight Center, 8800 Greenbelt Rd, Greenbelt, MD 20771, USA}



\begin{abstract}

We present the mid-IR (MIR) morphologies for 64 star-forming galaxies at $0.2<z<2.5$ with stellar mass $\rm{M_*>10^{9}~M_\odot}$ using JWST MIRI observations from the Cosmic Evolution Early Release Science survey (CEERS). The MIRI bands span the MIR (7.7--21~$\mu$m), enabling us to measure the effective radii ($R_{\rm{eff}}$) and S\'{e}rsic indexes of these SFGs at rest-frame 6.2 and 7.7 $\mu$m, which contains strong emission from Polycyclic aromatic hydrocarbon (PAH) features, a well-established tracer of star formation in galaxies. We define a ``PAH-band'' as the MIRI bandpass that contains these features at the redshift of the galaxy. We then compare the galaxy morphologies in the PAH-bands to those in rest-frame Near-UV (NUV) using HST ACS/F435W or ACS/F606W and optical/near-IR using HST WFC3/F160W imaging from UVCANDELS and CANDELS. The $R_{\rm{eff}}$ of galaxies in the PAH-band are slightly smaller ($\sim$10\%) than those in F160W for galaxies with $\rm{M_*\gtrsim10^{9.5}~M_\odot}$ at $z\leq1.2$, but the PAH-band and F160W have a similar fractions of light within 1 kpc. In contrast, the $R_{\rm{eff}}$ of galaxies in the NUV-band are larger, with lower fractions of light within 1 kpc compared to F160W for galaxies at $z\leq1.2$. Using the MIRI data to estimate the $\rm{SFR_{\rm{IR}}}$ surface density, we find the correlation between the $\rm{SFR_{\rm{IR}}}$ surface density and stellar mass has a steeper slope than that of the $\rm{SFR_{\rm{UV}}}$ surface density and stellar mass, suggesting more massive galaxies having increasing amounts of obscured fraction of star formation in their inner regions. This paper demonstrates how the high-angular resolution data from JWST/MIRI can reveal new information about the morphology of obscured-star formation.

\end{abstract}



\section{Introduction} \label{sec:intro}

Star formation and quenching mechanisms are the key to understanding the cosmic star formation history (e.g., \citealp{Madau2014}). 
However, it depends on a complex interplay of physical processes, including the rate at which gas accretes, cools, collapses and turns into stars, the effect of heavy elements and dust on cooling, and stellar and AGN feedback mechanisms.  
Studies on the galaxy structure up to $z \sim 2.5$ have converged to a coherent picture that the morphology of star forming galaxies (SFGs) are disk-dominated systems (with \citealt{Sersic63} indexes $n \simeq 1-2$), while quiescent galaxies are bulge-dominated (with $n \gtrsim 2$, e.g., \citealp{Shen2003, Wuyts2011, Weinzirl2011})\footnote{Throughout we will use \cite{Sersic63} indexes to model galaxy surface brightness profiles.  These are defined by $I(R) = I_e \exp\left\{ -b_n \left[ (R/R_\mathrm{eff})^{1/n} - 1\right] \right\}$ where $R_\mathrm{eff}$ is the effective radius, $n$ is the S\'ersic index, and $b_n$ is a constant chosen to ensure that the $R_e$ encloses 50\% of the total light.}. 
By measuring the galaxy size-mass distribution, it has been found that SFGs are on average larger than quiescent galaxies at all redshifts. 
Meanwhile, the slope of the size–mass relation for SFGs following $R_{\rm eff} \propto \rm M_*^{0.22}$, which is shallower than that for late-type galaxies ($R_{\rm eff} \propto \rm M_*^{0.75}$ e.g., \citealp{vanderWel2014, vanDokkum2015}). 
These studies quantify the relation between structure and star formation in galaxies from $z \sim 0$ up to $z \sim 3$. 
However, it remains unclear how the structural or size evolution proceeds.

Tracking spatially resolved star formation in galaxies will provide insight into this structural/size evolution, thus, their link to the dominant stellar buildup of galaxies and quenching mechanisms. 
Indeed, studies found that ongoing star formation traced by H$\alpha$ emission occurs in disks that are more extended than those occupied by existing stars in SFGs in the redshift range of $z\sim0.5 - 2.7$ \citep{Nelson2012,Tacchella2015, Nelson2016b, Wilman2020, Matharu2022}, while the extent of star formation and stellar disks are found to be the same in the local universe \citep{James2009, Fossati2013}. 
However, dust obscuration posts an immense challenge in uncovering star-formation activities via spatially resolved galaxy studies. 
As dust obscuration is more acute at shorter wavelengths, it preferentially impacts the UV and visible light emitted from stars.  Therefore, the possible presence of dust content could affect the interpretation of the observed light profile in UV/optical/NIR and nebular emission, particularly for massive SFGs at $\rm M_* \gtrsim 10^{10} M_\odot$, which are known to have more dust attenuation \citep{Whitaker2012,Nelson2016a, Tacchella2018}. 

Most of the aforementioned studies are limited by insufficient data to measure the spatially resolved dust profiles. Some studies have measured this using Balmer-line ratios (e.g., \ha/\hb) and rely on ``stacking'' to get sufficient signal-to-noise. For example, \citet{Nelson2016a} stacked the spatially-resolved dust attenuation maps using the Balmer decrements (H$\alpha$ and H$\beta$ emission) from the 3D-HST survey for galaxies with $\rm M_* > 10^{9} M_\odot$ and at $z\sim1$. They found that galaxies with $\rm M_* \gtrsim 10^{10} M_\odot$ have $\rm A_{H\alpha}\sim$ 2 mag of dust attenuation obscuring the star formation in their centers, while there is less dust attenuation obscuring the emission for lower-mass galaxies with $\rm M_* \lesssim 10^{10} M_\odot$ at all radii. 
In general, the H$\alpha$ emission should be more attenuated than stellar continuum, due to the addition attenuation on H$\alpha$ emission depending on the dust geometry (see review \citealp{Calzetti2001}). 
Therefore, without observations on the spatially-resolved dust, the total SFR and the spatial distribution of SFR and stellar remain unclear. 

Other studies have used size measurements of the dust emission in the far-IR (FIR) continuum measurements, e.g., 870 $\mu$m (rest-frame $\sim250 \mu$m  {at $z\sim2.5$}) obtained from ALMA.  These have generally found that the effective radii of galaxies in the rest-frame FIR are in general smaller than those of rest-frame optical or UV, revealed a more compact starburst region in these massive and/or dusty SFGs \citep{Hodge2015, Simpson2015, Chen2017, CalistroRivera2018, Gullberg2019, Hodge2019, Lang2019, Tadaki2017a, Tadaki2017b, Tadaki2020, Cheng2020, Gomez-Guijarro2022}. 
However, these studies focused on the bright sub-millimeter galaxies and/or massive galaxies with $\rm M_* \gtrsim 10^{11}\ M_\odot$ at high redshift $z > 1$.  
The spatially-resolved, or at least the extent of, obscured star formation of more common galaxies with $\rm M_* < 10^{10}\ M_\odot $  galaxies remain unknown. 

The Mid-Infrared Instrument (MIRI) on the James Webb Space Telescope (\jwst, \citealp{Gardner2006}) provide observations in the mid-infrared (MIR) region of the electromagnetic spectrum spanning the wavelength range of 7.7--21~$\mu$m.
Particularly, the MIRI data covers the Polycyclic Aromatic Hydrocarbons (PAH) features at 7.7$\mu$m up to $z\sim1.7$ and 6.2$\mu$m up to $z\sim2.5$, both of them tracing photodissociation regions (PDRs) associated with H II regions (e.g., \citealp{Calzetti2007}). 
In addition, the high spatial resolution of MIRI (FWHM = $0\farcs3$ at 10$\mu$m, corresponding to 2.4 kpc at $z\sim1$) enables studies of resolved morphological structures in distant galaxies for the first time. 

In this paper, we use the MIRI data taken as part of the Cosmic Evolution Early Release Science survey (CEERS, Finkelstein et al. \textit{in prep.}, Yang et al. \textit{in prep.}) to measure the morphology of galaxies in rest-frame MIR to trace the obscured star formation region for galaxies down to stellar mass of $\rm M_* \sim 10^{9} M_\odot$ in the redshift range of $0.2 < z < 2.5$. We compare these data to data from the \textit{Hubble Space Telescope} (\hst) covering the rest-frame NUV for these galaxies from the UVCANDELS survey which traces their unobscured star formation. 
Here we focus on a comparison of the NUV and MIR morphologies as this allows us to make a more complete picture of the morphology of star formation. Meanwhile, we anchor these results against the \hst/WFC3 F160W imaging from CANDELS covering the rest-frame optical/NIR for these galaxies, which traces the existing stars. 
{Magnelli} et al., (\textit{in prep.}) will perform a more detailed analysis of the morphologies of thermal dust using the the full CEERS MIRI data (4 pointings) and {focus on more massive galaxies ($\rm M_* \gtrsim 10^{10} M_\odot$).  They will also compare these to other results of the} dust sizes derived from far-IR imaging from ALMA {available in other fields}. 

This paper is laid out as follows. 
Section~\ref{sec:data} provides an overview of Mid-IR, optical/NIR, and UV data, and multi-wavelength catalogs. 
We describe the bandpasses selection, sample selection, morphology measurements, {SED fitting} and surface density of stellar mass and star formation rate (SFR) in Section~\ref{sec:method}. 
In Section~\ref{sec:result}, we presents our results. 
We discuss the robustness of our results and physical implications in Section~\ref{sec:discussion}. 
We conclude with a summary in Section~\ref{sec:summary}. 
Throughout this paper, all magnitudes, including those in the IR, are presented in the AB system \citep{Oke1983, Fukugita1996}. We adopt a standard concordance Lambda cold dark matter
($\Lambda$CDM) cosmology with $H_0$ = 70 km s$^{-1}$, $\Omega_{\rm \Lambda}$ = 0.70, and $\Omega_{\rm M}$ = 0.30.

\section{Survey and Data} \label{sec:data}

\subsection{CEERS} \label{sec:ceers}
The Cosmic Evolution Early Release Science survey (CEERS, Finkelstein et al., \textit{in prep.}) is an Early Release Science (ERS) program (Proposal ID \#1345) that will cover a total $\sim100$ sq. arcmin of the Extended Groth Strip field (EGS, \citealp{Davis2007}). This field is one of the five legacy fields of the Cosmic Assembly Near-IR Deep Extragalactic Legacy Survey (CANDELS; \citealp{Koekemoer2011, Grogin2011}) and partially covered by the 3D-HST Treasury Survey \citep{Skelton2014, Momcheva2016}. 
CEERS will obtain observations in several different modes with JWST, including NIRCam imaging \citep{Bagley2022}, MIRI imaging (Yang et al. in prep) and NIRSpec multi-object spectroscopic observations (see Arrabal Haro et al. \textit{in prep.}). 

The MIRI imaging of CEERS includes seven filters (F560W, F770W, F1000W, F1280W, F1500W, F1800W, and F2100W). 
The first set of CEERS observations were taken on 21 June 2022 in four pointings, named as CEERS1, CEERS2, CEERS3, and CEERS6 (see Table 1 in \citealt{Bagley2022}).  These included MIRI imaging with NIRCam in parallel. 
In this paper, we focus on the MIRI imaging in the CEERS1 and CEERS2, which  received coverage in the filters covering longer wavelengths (F770W, F1000W, F1280W, F1500W, F1800W, and F2100W), covering important rest-frame mid-infrared emission features for galaxies at high redshift (the CEERS3 and 6 pointings cover the shorter-wavelength MIRI filters, F560W and F770W, see \citealp{Papovich2022}). In particular, the MIRI observations covered the PAH feature at 7.7\ $\mu$m for galaxies up to $z = 2$). 

\subsection{Mid-IR imaging and catalog} \label{sec:MIRI_data}

A description of the properties of the MIRI data and the reduction of these data will appear elsewhere (Yang et al. \textit{in prep.}). 
We summarize the steps here.  The data were processed using \jwst\
\texttt{calibration pipeline} (v1.7.2) using mostly the default
parameters for stage 1 and 2.  We then modelled the background by taking the median of all the other images taken in the same bandpass but at different fields and/or dither positions.  We then subtracted this background from each image.  
The astromety are corrected by matching to the CANDELS \hst\ imaging (\citealt{Koekemoer2011, Bagley2022}\footnote{\url{https://ceers.github.io/releases.html}}\label{footnote:ceers_hdr}) prior to processing the images with stage 3 of the pipeline.  This produced the final science images, weight maps, and uncertainty images (the latter include an estimate for correlated pixel noise; see Yang et al. \textit{in prep.}) with a pixel scale of $0\farcs09$, registered to the \hst/CANDELS v1.9 WFC3 images.  

The MIRI photometry is measured for sources detected in the original CANDELS \hst/WFC3 catalog from \citet{Stefanon2017}. This is appropriate as our primary interest here is in studying the MIRI morphologies and comparing them to the \hst\ rest-frame NUV and optical data. To measure fluxes, we use \texttt{TPHOT} \citep{Merlin2016} which uses an image with higher angular resolution (in this case the \hst/WFC3 F160W detection image, which has a Point Spread Function (PSF) with FWHM of $\simeq$0.2\arcsec) as a prior for photometry in images with lower angular resolution (in this case the MIRI images, which have a PSF with FWHM of $\simeq$0.2--0.5\arcsec). 
The PSF for each MIRI band are constructed using WebbPSF.  
We then constructed kernels to match the PSF of the F160W data to the MIRI bands.  With these, we performed source photometry with \texttt{TPHOT}. This provides MIRI flux densities and uncertainties for each source in the CANDELS \hst/F160W catalog, which we use as our MIRI catalog. 

\subsection{UV imaging and catalog} \label{sec:UV_data}

We adopted the HST WFC3/F275W and ACS/F435W imaging as part of the Ultraviolet Imaging of the Cosmic Assembly Near-infrared Deep Extragalactic Legacy Survey Fields (UVCANDELS) Hubble Treasury program (GO-15647, PI: H. Teplitz), which covers four of the five premier CANDELS fields (GOODS-N, GOODS-S, COSMOS and EGS).
The primary WFC3/F275W imaging reached $m \leq 27$~AB mag for compact galaxies (SFR$\sim$0.2 M$_\odot$/yr at $z=1$), and the coordinated parallel ACS/F435W imaging reached $m \leq 28$~AB mag.
We adopt two methodologies to measure the F275W and F435W fluxes, consistent with the previous measurements at other wavelengths (Wang et al. \textit{in prep}).
First, we adopt the conventional hot+cold method based on object near-infrared isophotes and PSF matching following the CANDELS methodology, as in \cite{Stefanon2017}.
We also derive the UV-optimized aperture photometry method based on object optical isophotes aperture, following the work done in the Hubble Ultra-Deep Field UV analysis \citep{Teplitz2013, Rafelski2015}.  By using smaller optical apertures without degradation of the image quality, our UV-optimized aperture photometry method reaches the expected 5-$\sigma$ point-source depth of 27 AB mag in F275W, deeper by $\sim$1 AB mag than the depth reached by the conventional hot+cold method. 
On average, our UV-optimized photometry yields a factor of 1.5$\times$ increase in signal-to-noise ratio in the F275W band, with higher increase in brighter extended objects. Henceforth, we take the F275W and F435W photometry obtained from the latter method as our default measurements, which complement the pre-existing CANDELS photometric catalog presented in \cite{Stefanon2017}.  

\subsection{Optical/Near-IR Imaging {and Catalog}}

For the analysis in this paper, we adopt the multi-wavelength photometric catalog from HST observations from \citet{Stefanon2017} (``S17 catalog'' hereafter) as our primary catalog for our study here. 
This catalog provide measurements of the photometric redshifts and stellar population parameters of galaxies in the EGS field, using broad/median-band UV/NIR data spanning from 0.4 to 8$\mu$m, taken by six different instruments, including Canada France Hawaii Telescope (CFHT)/MegaCam, NEWFIRM/NEWFIRM, CFHT/WIRCAM, HST/ACS, HST/WFC3, and Spitzer/IRAC. 
The  multi-band photometric data were independently analyzed by 10 different {groups, each one using a different set of code and/or SED templates}, including FAST \citep{Kriek2009}, HyperZ \citep{Bolzonella2000}, Le Phare \citep{Ilbert2006}, WikZ \citep{Wiklind2008}, SpeedyMC \citep{Acquaviva2012}, and other available codes \citep{Fontana2000, Lee2010}. 
The final photometric redshifts are the median of these 10 photometric redshifts. 
The final stellar masses are computed as the median of the results from the six sets which adopted an exponentially declining SFH and the \citet{Chabrier2003} IMF, without considering nebular emission contamination {(M2, M6, M10, M11, M13 and M14 in Tab. 4 of \citealp{Stefanon2017})}. 
{Note that the photometric redshifts and stellar mass do not incorporate the MIRI data. However, we do not observe significant differences between stellar mass from S17 and stellar mass from the CIGALE SED fitting which includes MIR and FUV data (see Section \ref{sec:cigale}). Therefore, for this analysis, }
we adopt the final photometric redshifts and stellar masses presented in the ``zbest'' and ``M\_med'' columns in the S17 catalog {and their associated uncertainties. }

We then remove AGN using the ``AGNflag'' in S17, which flags AGN by cross-matching to sources in the Chandra X-ray data from the AEGIS project (AEGIS-X Wide, \citealp{Nandra2005, Laird2009} and AEGIS-XD, \citealp{Nandra2015}).  {We also consider galaxies with AGN that are missed by the X-ray data. We search for these using the \cigale\ SED fitting, and remove 6 additional galaxies where \cigale\ finds an AGN component could contribute more than 10\% to the total IR luminosity (see Section~\ref{sec:sample} and Section~\ref{sec:cigale}).} 

\subsection{{Additional Mid-IR/Far-IR Imaging and Catalog}} \label{sec:mips}

{In addition, we adopt the IRAC 3.6+4.5 \micron\ selected multi-wavelength catalog that contains Spitzer/MIPS 24 \micron\ and 70 \micron\ fluxes from \citet{Barro2011} (``B11 catalog'' hereafter). 
These MIPS data are obtained as part of the Guaranteed Time Observations (GTO, PI: Fazio) and the Far-Infrared Deep Extragalactic Legacy Survey (FIDEL). The 5$\sigma$ limiting magnitude of MIPS 24 \micron\ and 70 \micron\ are 60 $\mu$Jy and 3.5 mJy, respectively. 
We then cross-match the S17 catalog to the B11 catalog using a 1$^{\prime\prime}$ radius. }

\section{Sample Selection and Methods}\label{sec:method}

\subsection{Bandpass Selection} \label{sec:bands}

\begin{figure}
    \centering
    \includegraphics[width=\columnwidth]{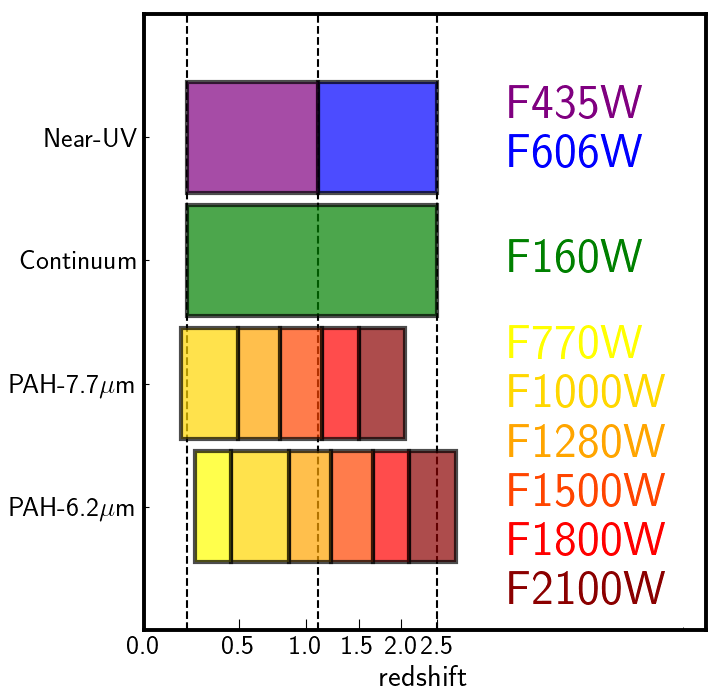}
    \caption{Selection of \jwst/MIRI and \hst\ bandpasses as PAH-, NUV-bands and F160W for morphological measurements {of galaxies at $0.2 < z < 2.5$}. For each galaxy, we select the PAH-band(s) from one or two MIRI bands that contain the rest-frame 7.7~$\mu$m and 6.2~$\mu$m PAH features. The yellow to red color shaded regions represent the redshift coverage of MIRI bandpasses. We adopt the \hst/F160W for all galaxies, as it covers their rest-frame optical/NIR regime, corresponding to stellar continuum. We use either the \hst/WFC3 F435W bandpass (for galaxies at $z < 1.1$ [purple shaded]) or the \hst/ACS F606W bandpass ($z > 1.1$ [blue shaded]) as the NUV-band that contain the rest-frame NUV of these galaxies, which traces the unobscured star formation. {Three relevant redshifts ($z$ = 0.2, 1.1, 2.5) in the band selection are marked as vertical dashed lines.}}     \label{fig:band}
\end{figure}

We take advantage of the fact that the \jwst/MIRI, \hst/F160W, F606W, and F435W imaging are dominated by emission that originates from different regions in distant galaxies (i.e., massive stars and long-lived stars). 
Here, we select bands specifically to best isolate them. 
PAH emission arises from photodissociation regions (PDR) around \ion{H}{2} regions of young, massive stars.  Thus, PAH emission traces star formation. Because the PAHs emit at the mid-IR wavelengths, they are much less affected by dust attenuation, and therefore probe more obscured star formation. 
Previous work has shown that the integrated PAH luminosity -- SFR relation has been calibrated for galaxies up to $z\sim0.4$ \citep{Shipley2016, Xie2019}. 
The total PAH emission can contribute 10-20\% of the total IR luminosity, and the 7.7~$\mu$m PAH feature, the strongest PAH feature, may contribute $\sim$50\% of the total PAH emission (e.g., \citealp{Smith2007, Wu2010, Shipley2013}). 
The next relative strong PAH feature is at 6.2~$\mu$m, which has the benefits of being relatively isolated with little contamination from nearby features and observable in galaxies at higher redshift up to $z\sim2.5$ with MIRI. 
In our study, we select one or two MIRI bandpasses that contain the rest-frame 6.2 and/or 7.7~\micron\ PAH features for each galaxy, as shown in Fig.~\ref{fig:band}. We refer to them as the ``PAH-bands''. We measure the morphology of galaxies in these PAH-bands and interpret them as the morphology of obscured star formation regions. 

\begin{figure*}
\includegraphics[width=0.9\textwidth]{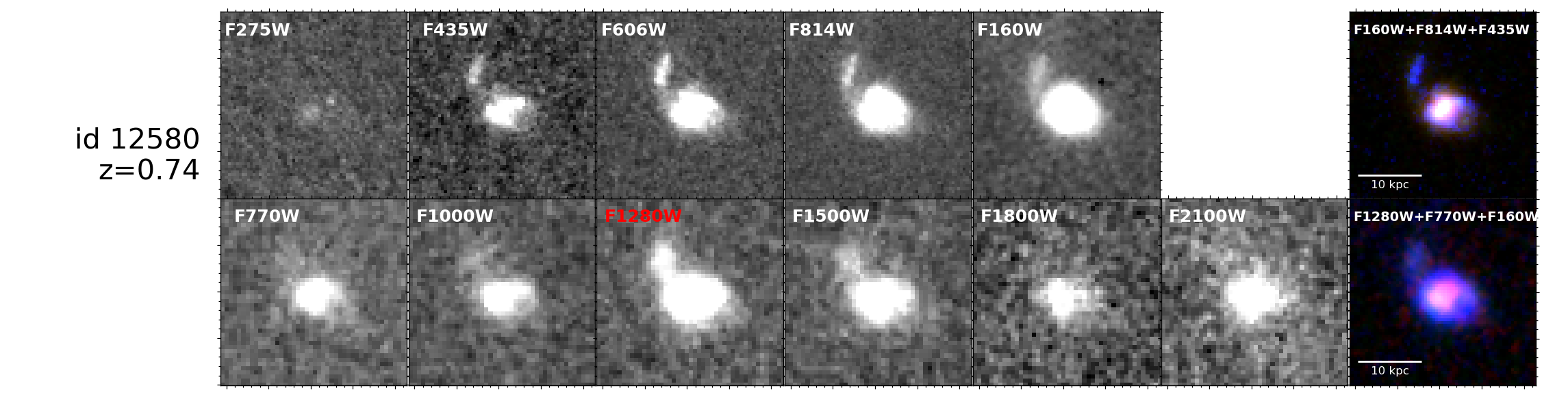}
\includegraphics[width=0.9\textwidth]{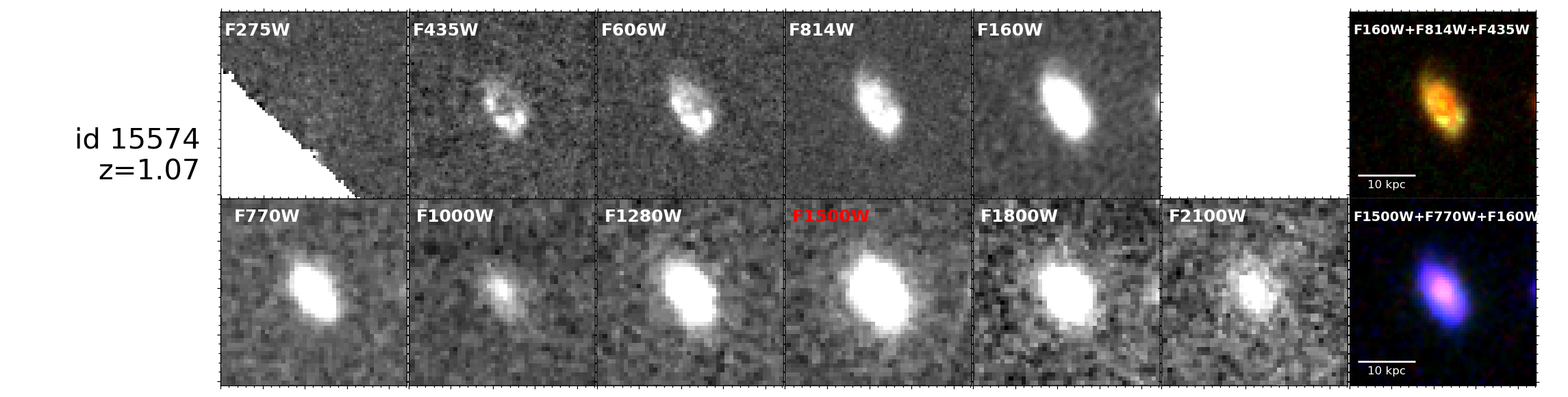}
\includegraphics[width=0.9\textwidth]{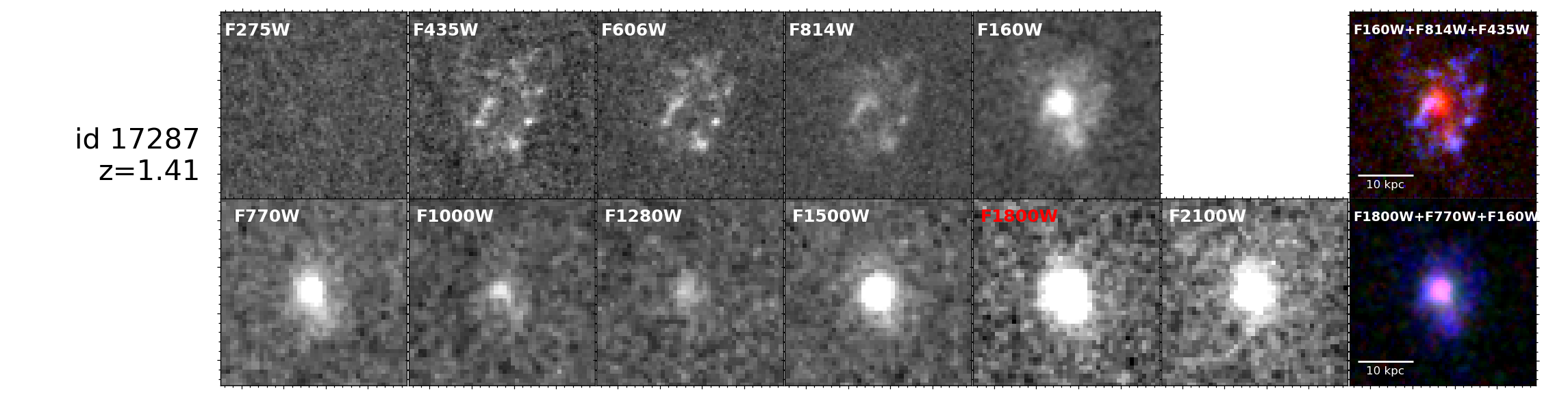}
\includegraphics[width=0.9\textwidth]{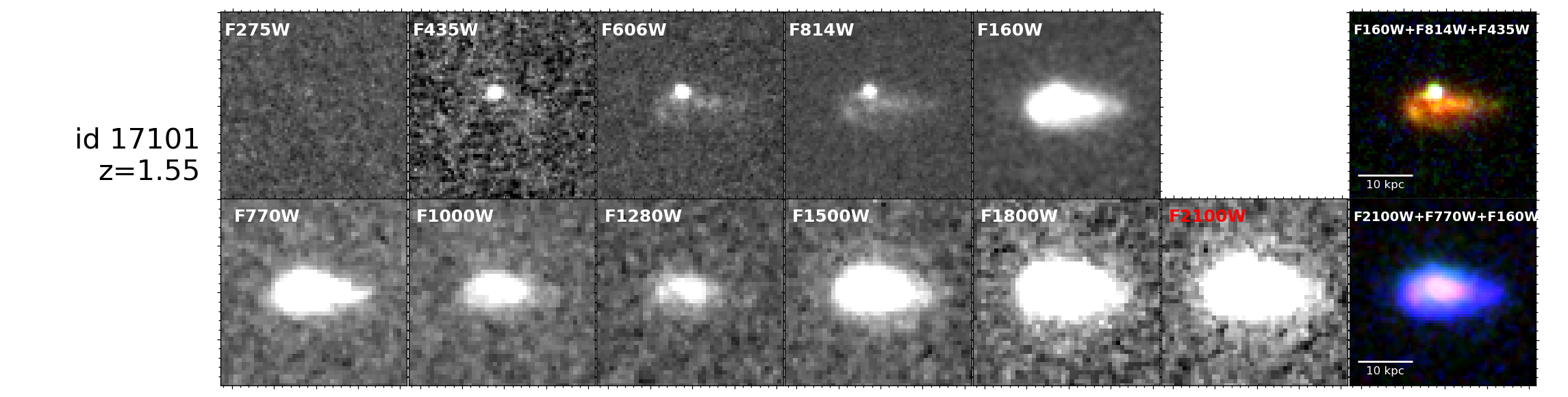}
\includegraphics[width=0.9\textwidth]{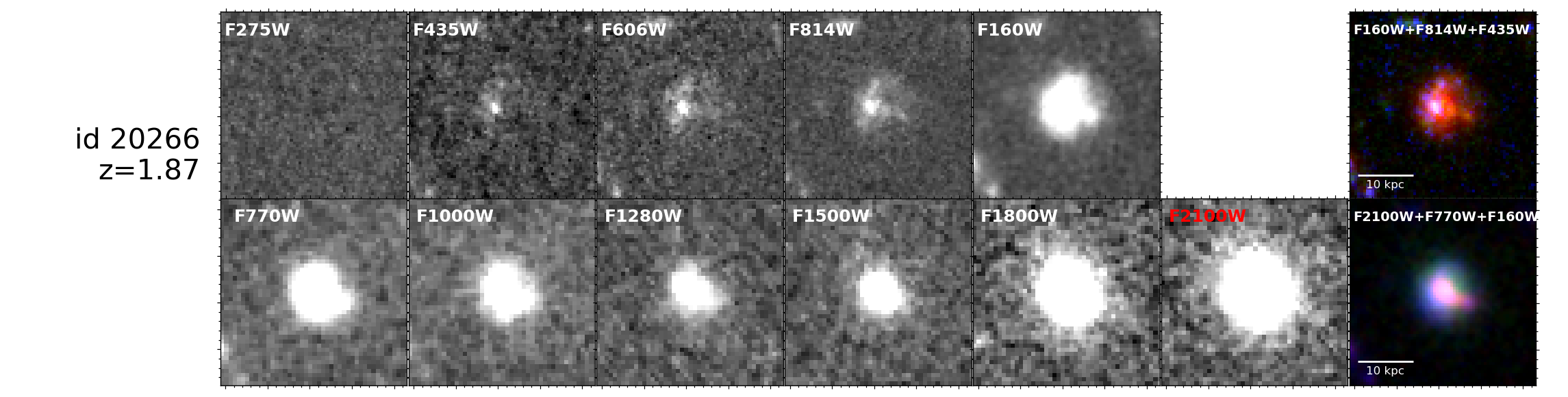}
\caption{Example of the postage stamps ($2^{\prime\prime}\times2^{\prime\prime}$) centered on the selected galaxies in all available \hst\ and \jwst\ images, ordered by increasing redshift. In the right-most panels, we show false color image using HST F435W+F814W+F160W (\textit{top}) and F160W/HST+F770W/JWST+PAH-band at rest-frame 7.7~$\mu$m from MIRI/JWST (\textit{bottom}) as blue+green+red false color, respectively. For each galaxy, the PAH band at rest-frame 7.7~$\mu$m is marked by red label. The ID and photometric redshift from \citet{Stefanon2017} are marked on the left.
The complete figure set (64 images) is available in the online journal.} 
\label{fig:cutouts}

\figsetstart
\figsetnum{2}
\figsettitle{The false color images (HST F435W+F814W+F160W and F160W/HST+F770W/JWST+PAH-band at rest-frame 7.7~$\mu$m from MIRI/JWST) for all galaxies in the final sample}

\figsetend
\end{figure*}

To trace the profile of unobscured star formation, we use either 
 \hst/F435W or \hst/F606W for galaxies at $z \leq 1.1$ or $z > 1.1$, respectively, which covers the rest-frame Near-UV (NUV) regime of galaxies (named as the ``NUV-band''). 
Note that, ideally, the rest-frame Far-UV should be used, which tie more closely to massive stars. However, due to the shallower depth of the F275W imaging and the complex Far-UV structure, most of galaxies have lower signal-to-noise (S/N), and/or are resolved into a few star-forming clumps \citep{Mehta2022}. 
Thus, we choose to measure the NUV morphology. 
For the profile of stellar continuum, we adopted the \hst\ WFC3/F160W from CANDELS, which covers the rest-frame optical regime of galaxies at $z\sim1$. 
The bandpass selection of the PAH-, NUV-bands and F160W are summarized in Fig.~\ref{fig:band}. 
Fig. \ref{fig:cutouts} displays the \hst\ and \jwst\ data and false color images for selected galaxies. {The complete figure set of the false color images for galaxies in the final sample  (64 images) is available in the online journal. }

\subsection{Sample Selection} \label{sec:sample}

\begin{figure*}
    \centering
    \includegraphics[width=\textwidth]{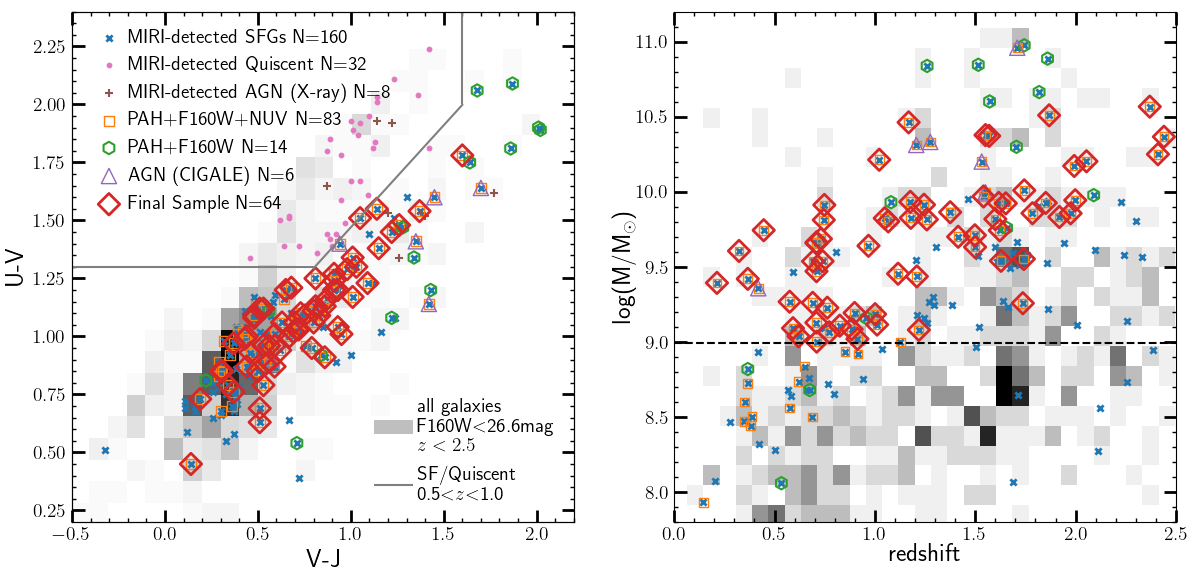}
    \caption{{A $UVJ$ color -- color (\textit{left}) and stellar mass -- redshift (\textit{right}) phase diagram of the final sample}. In the each panel, MIRI-detected SFGs (blue cross), $UVJ$-selected quiescent galaxies (pink dots), AGN identified by X-ray detections (brown pluses) are overlaid on the 2D histogram of all photometric galaxies at $z < 2.5$. Open symbols mark those MIRI-detected SFGs with successful \galfit\ fits in all three of the PAH-, NUV-bands, and F160W (orange), and only in PAH-band and F160W (green). Additional AGN candidates identified by \cigale\ are marked by purple triangles. The final sample are marked by red open diamonds. The solid grey lines in the left panel show the separation for galaxies at $0.5<z<1.0$ applied to our sample to select SFGs \citep{Williams2009}. The dashed back line in the right panel indicates the stellar mass selection limit.  }
    \label{fig:prop}
\end{figure*}

To construct our sample, we start with the MIRI fluxes for sources in the S17 catalog. We incorporated spectroscopic redshifts ($z_{spec}$) from various spectroscopic surveys, including the DEEP2 Galaxy Redshift Survey \citep{Newman2013}, the DEEP3 Galaxy Redshift Survey \citep{Copper2012, Zhou2019}, the MOSFIRE Deep Evolution Field (MOSDEF) survey \citep{Kriek2015}, and the Complete Calibration of the Color-Redshift Relation (C3R2) survey \citep{Masters2017}. 
We selected sources with F160W magnitude $<26.6$ AB magnitude (this is the 90\% completeness for point sources), photometric redshift $z_{phot} \leq 2.5$ ($z_{spec}$ when available), and without any AGN flags \citep{Stefanon2017}. The redshift cut is based on the redshift coverage of MIRI bandpasses for 6.2~$\mu$m PAH-band (see Fig~\ref{fig:band}. 
We then selected galaxies with detection $>5\sigma$ in their selected MIRI bands (see Section~\ref{sec:bands}). 
To select SFGs, we adopted the $UVJ$ color -- color separation as function of redshift from \citet{Williams2009} (see Fig.~\ref{fig:prop}). 
These gives us a sample of total 161 MIRI-detected SFGs. 

For our study we need to derive measurements on the morphological parameters in multiple bandpasses, the NUV-band, F160W, and PAH-band(s). This requires sufficient signal-to-noise in each bandpass. We therefore model the morphological parameters in all bands for the sample of 161 MIRI-detected SFGs using \galfit\ (see more details in section \ref{sec:size}) and we refine our sample to include only objects for which the \galfit\ model fit converges in all of NUV-, PAH-band and F160W. 

\begin{figure}
    \centering
    \includegraphics[width=\columnwidth]{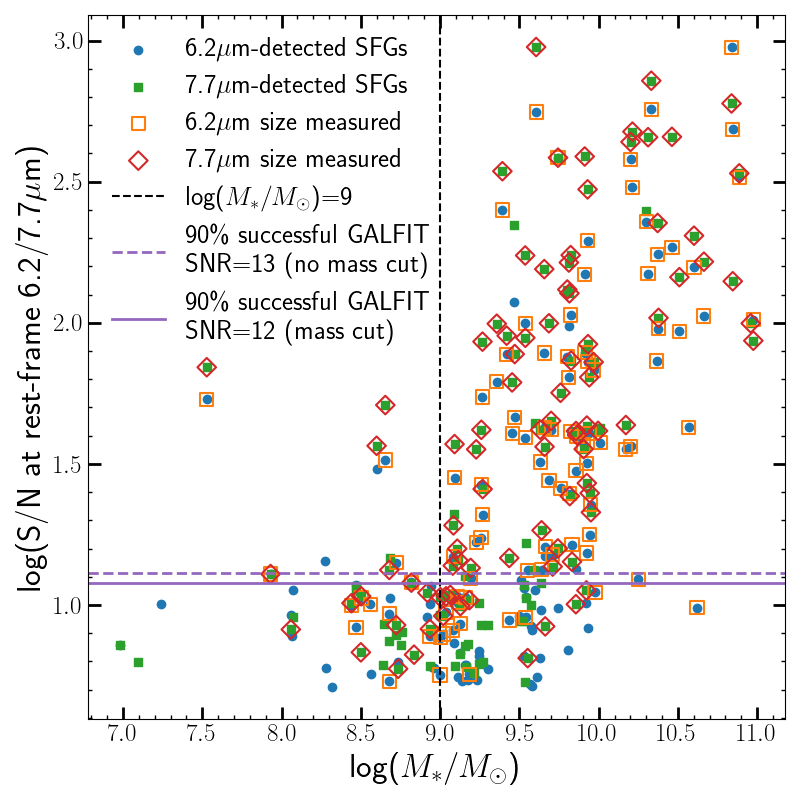}
    \caption{The signal-to-noise ratio of the \jwst/MIRI detection in the PAH-band (rest-frame 6.2/7.7~$\mu$m)) as function of stellar mass for the MIRI-detected SFGs. The blue dots and green squares correspond to MIRI-detected SFGs selected with rest-frame 6.2~$\mu$m and 7.7~$\mu$m MIRI flux $>5\sigma$, respectively. The open red and orange symbols marked those which have secure morphology measurements using \galfit. The black vertical line marks the stellar mass cut at $M_*\geq10^9 M_\odot$. The purple dash and solid line show the SNR of MIRI fluxes that 90\% of galaxies have successful \galfit\ fits. }
    \label{fig:snr}
\end{figure}

We first model the morphologies of 161 MIRI-detected SFGs in the PAH-bands (the one or two MIRI imaging that contain the rest-frame 6.2 and 7.7~\micron\ PAH features) using \galfit, discussed in Section~\ref{sec:size}. We obtain 106 with successful morphological measurements in the PAH-band(s).  The remaining sources where \galfit\ fails to converge have {in general} lower signal-to-noise ratios (SNRs).  This limits our sample of SFGs in SNR to $\gtrsim 12-13$ (see Fig.~\ref{fig:snr}). 
For galaxies which have successful morphological measurements in both of the MIRI bands (i.e., the two that contain the 6.2 and 7.7~\micron\ PAH emission), we adopt the fits from the band with higher SNR in flux density.
For the majority of our galaxies (76 out of 106 MIRI-detected SFGs), we adopt the morphological measurements using the PAH-band contains rest-frame 7.7$\mu$m PAH feature.  This is primarily because the 7.7$\mu$m PAH feature dominates the total PAH emission (e.g., \citealt{Smith2007}). 
We note that choosing to adopt PAH morphologies measurements at either rest-frame 6.2 or 7.7~$\mu$m does not affect any result shown in this paper. 

We then model the morphologies of these 106 MIRI-detected SFGs in the NUV-band (either the \hst\ WFC3/F435W or ACS/F606W) and F160W images using \galfit (see Section~\ref{sec:size}). Of these, 83 SFGs have successful \galfit\ model fits measured in all three bands (the NUV-, PAH-bands and F160W). 
There are 14 galaxies that have morphologies where \galfit\ is successful in only F160W and the PAH-band, but fails in the NUV-band (again, because the SNR is too low in the latter).  
Of these galaxies where \galfit\ is unsuccessful in the NUV-band, they are located in the dusty region of the $UVJ$ color-color diagram and are more massive with median stellar mass of $\rm \langle M_* \rangle = 10^{10.3} M_\odot$ than galaxies having secure sizes measured in all three bands ($\langle M_*\rangle = 10^{9.7} M_\odot$, see Fig. \ref{fig:prop}). 
This is most likely due to the higher dust attenuation in these galaxies absorbing most of NUV light. 
{Due to the small number of these galaxies, we exclude them in the analyses of this paper. However, we defer an analysis of these galaxies to a future study (Magnelli et al., \textit{in prep.}). }
There are additional 7 galaxies where the \galfit\ model is successful in the NUV- and PAH bands, but fails in F160W, and 2 more galaxies where \galfit\ fails in both the F160W and NUV band. { Due to the small number of these galaxies and the primary focus of this paper, we also exclude them in the analyses of this paper. }

We impose a stellar mass selection of log($M_*/M_\odot$) $\geq$ 9, where the sample is highly complete (S17 reports that log($M_*/M_\odot$) $\geq$ 9 is the 90\% completeness of point-source detection at $z\sim1$ assuming a passively evolving simple stellar population (SSP) model \citealp{Bruzual2003} with $A_{\rm V}$ = 3 mag).
This stellar mass selection effectively removes galaxies with low SNR in the PAH-band(s) and limits the sample to object where more than $\sim$10\% of the sample has a successful \galfit\ model fit (Fig.~\ref{fig:snr}).  
\cite{Matharu2022} found that size measurements in CANDELS-like \textit{HST} imaging are less reliable for galaxies at these redshifts with $\rm M_* \lesssim 10^{9} M_\odot$ (primarily this is because the sizes of the galaxies are small and approaching the resolution limit of the \hst\ WFC3 image.). We therefore apply this mass limit to our study here.  

{Finally, we remove 6 AGN candidates identified as $f_{\rm AGN} \geq 0.1$ from CIGALE SED fitting (see Section \ref{sec:cigale}).} 
Our final sample includes {64} MIRI-detected SFGs at $0.2 < z < 2.5$ with morphological measurements in the NUV-, PAH-bands and F160W. This includes 14 and {48} galaxies with their PAH morphologies measured at rest-frame 6.2$\mu$m and 7.7$\mu$m, respectively. {There are 26 out of 64 have spectroscopic redshifts. The median uncertainties on photometric redshift of our final sample is 0.12. }
A summary of the sample selection and number of galaxies is listed in Table~\ref{tab:sample}.  
The median redshift and stellar mass and their 16th/84th percentiles of our final MIRI-detected SFGs are listed in Tab.~\ref{tab:props}. 
{The scatter plot of stellar mass and redshift for MIRI-detected SFGs, those with secured morphologies measured in all three bands and the final sample are shown in Fig.~\ref{fig:prop}. } 

We note a potential bias of the sample selection toward SFGs with moderate dust and against very dusty SFGs because of excluding NUV-undetected sources.
Our goal is to focus on the comparison between NUV and PAH morphologies. Magnelli. et al \textit{in prep.} will focus on the morphology studies from the MIRI data that include objects undetected in the NUV-band.

%
%The median redshift and stellar mass and their 16th/84th percentiles of our final MIRI-detected SFGs are $1.15_{-0.44}^{+0.59}$ and $9.69_{-0.51}^{+0.47} \rm M_\odot$, respectively. 
%

\begin{deluxetable}{l c}
\tablecolumns{2}
\tablecaption{Summary of sample selection\label{tab:sample}}
\tablehead{ \colhead{Selection Criteria} & \colhead{Num. of Gals.} \\
\colhead{(1)} & \colhead{(2)}}
\startdata
{Sources in the parent MIRI catalog } & {964} \\[5pt]
$m$(F160W) $\leq$ 26.6 AB and $z_{phot}(z_{spec}) \leq 2.5$ & 683 \\[5pt]
AGN flag = 0 & 675 \\[5pt]
Rest-frame 6.2 or 7.7~$\mu$m PAH-band & \multirow{2}{*}{192}  \\
 SNR $> 5$ & \\ [5pt]%(180/156)
$UVJ$-selected, MIRI-detected SFGs\tablenotemark{$\dag$} & 161 \\ [5pt]% 
%\hline
%\cutinhead{Successful GALFIT model fits in:}
Successful \galfit\ model fit in  & \multirow{2}{*}{106 (76/30)} \\
MIRI PAH-band(s)  & \\[5pt]
Successful \galfit\ model fit in   & \multirow{2}{*}{83 (65/18)} \\
PAH-band(s), F160W \& NUV-band& \\[5pt]
%F160W \& PAH  & 13 \\
%NUV \& PAH & 7 \\
%only PAH  & 2 \\
%\hline
%\cutinhead{Final Sample:}
%Final Sample with log($M_*/M_\odot) \geq 9$   & 70 (56/14)\\ 
log($M_*/M_\odot) \geq 9$  & 70 (56/14)\\[5pt]
Final Sample with $f_{\rm AGN} < 0.1$ from \textsc{cigale} & 64 (50/14)\\ 
\enddata
\tablecomments{(1) The selection criteria and sample name, note the selection criteria are cumulative; (2) The total number of galaxies. Numbers in the parentheses are number of galaxies use morphology measured in rest-frame 7.7$\mu$m or 6.2$\mu$m, respectively. }
\tablenotetext{$\dag$}{We remove ``quiescent'' galaxies following the rest-frame color selection of \citet{Williams2009}.}
\end{deluxetable}

\begin{figure*}
\centering
\gridline{
\includegraphics[width=0.4\textwidth]{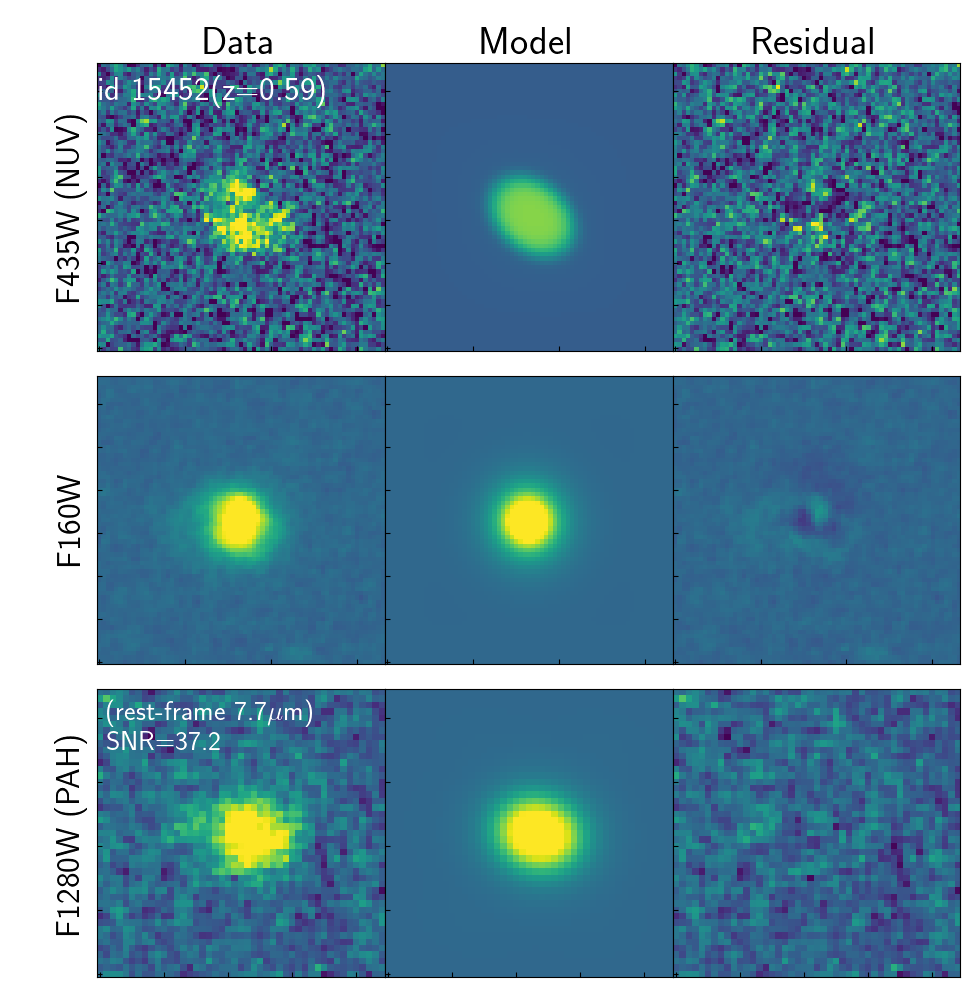}
\includegraphics[width=0.4\textwidth]{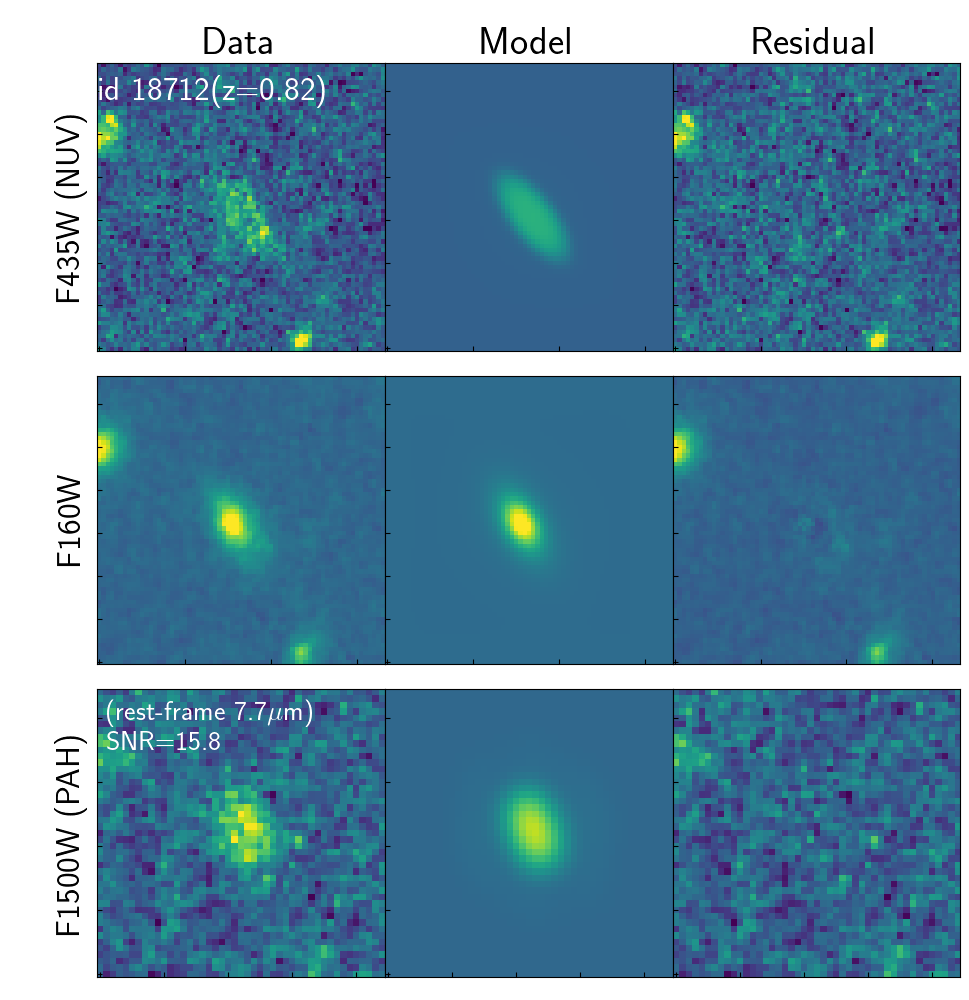}
}
\vspace{-5mm}
\gridline{
\includegraphics[width=0.4\textwidth]{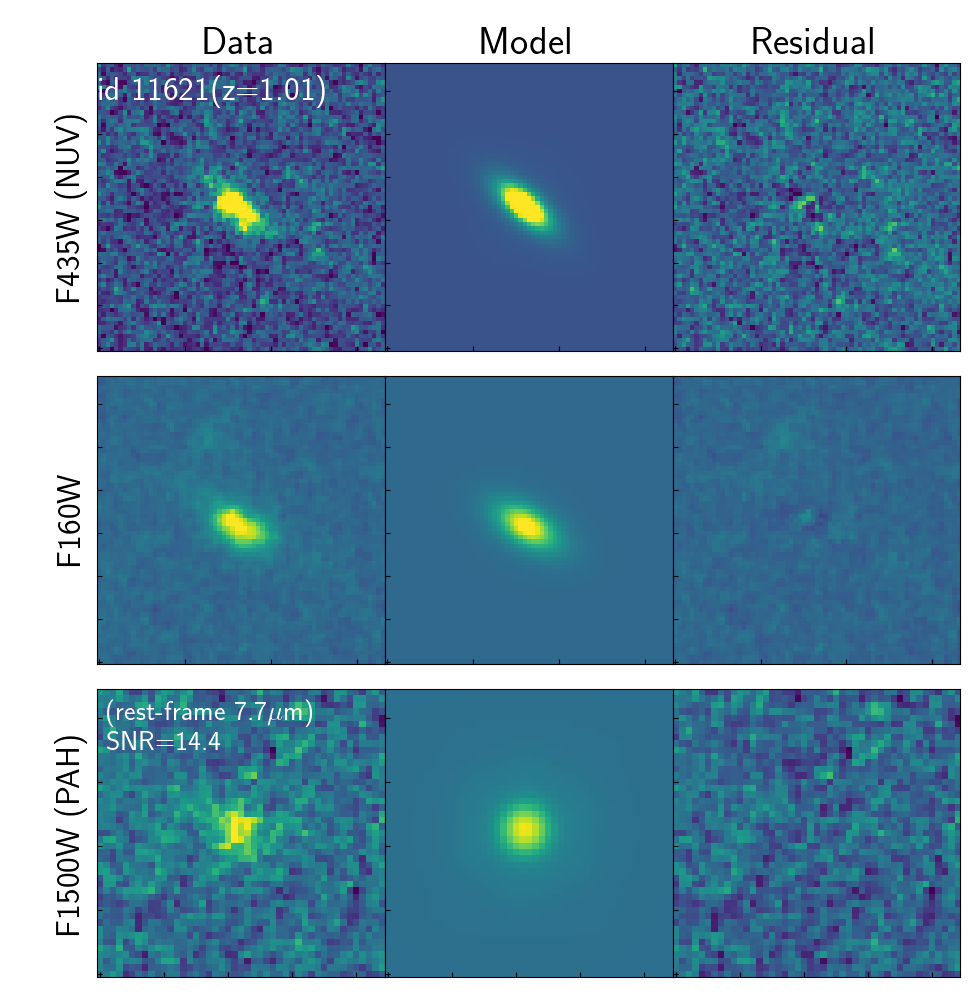}
\includegraphics[width=0.4\textwidth]{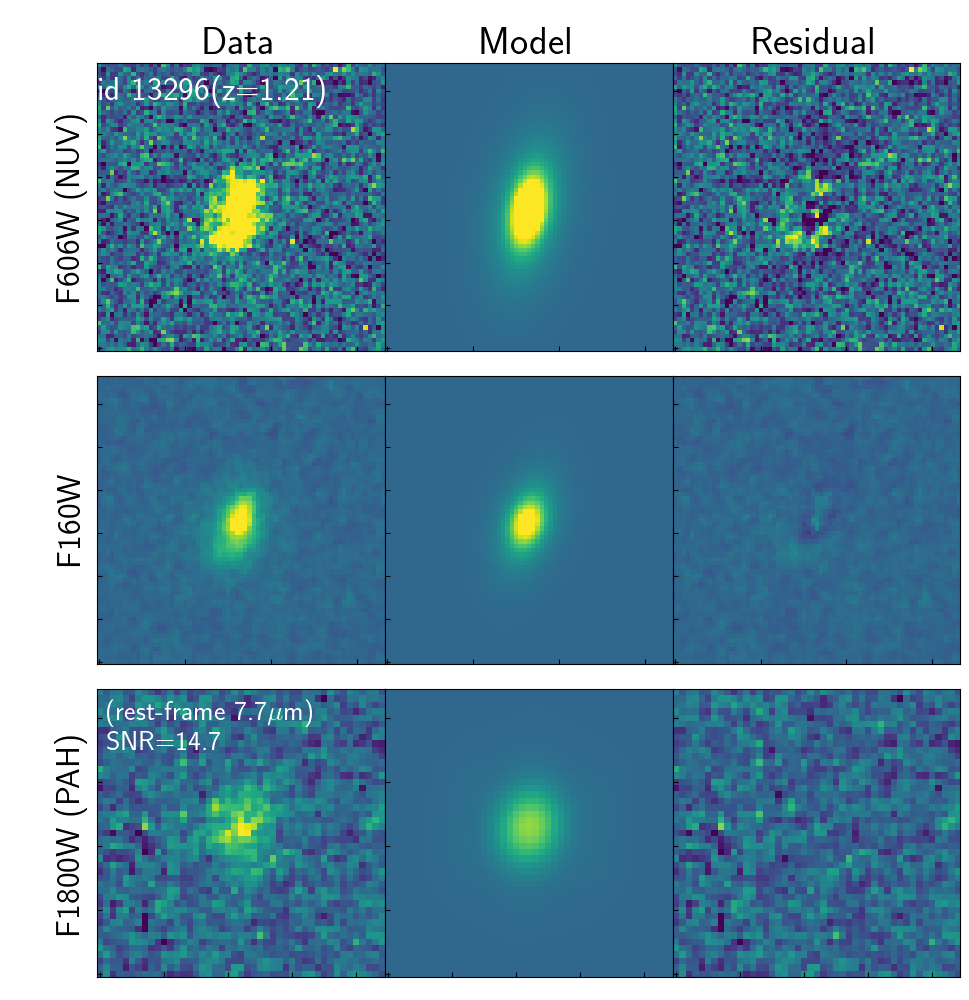} 
}
\vspace{-5mm}
\gridline{
\includegraphics[width=0.4\textwidth]{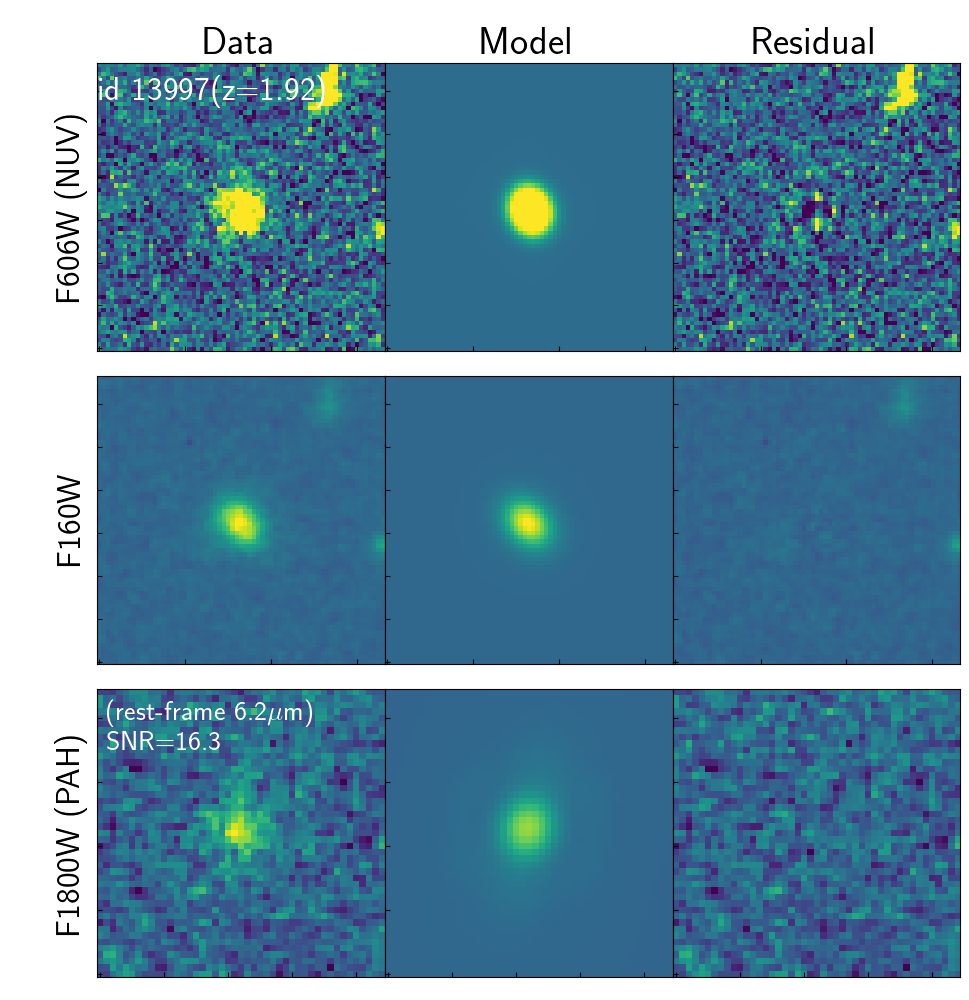}
\includegraphics[width=0.4\textwidth]{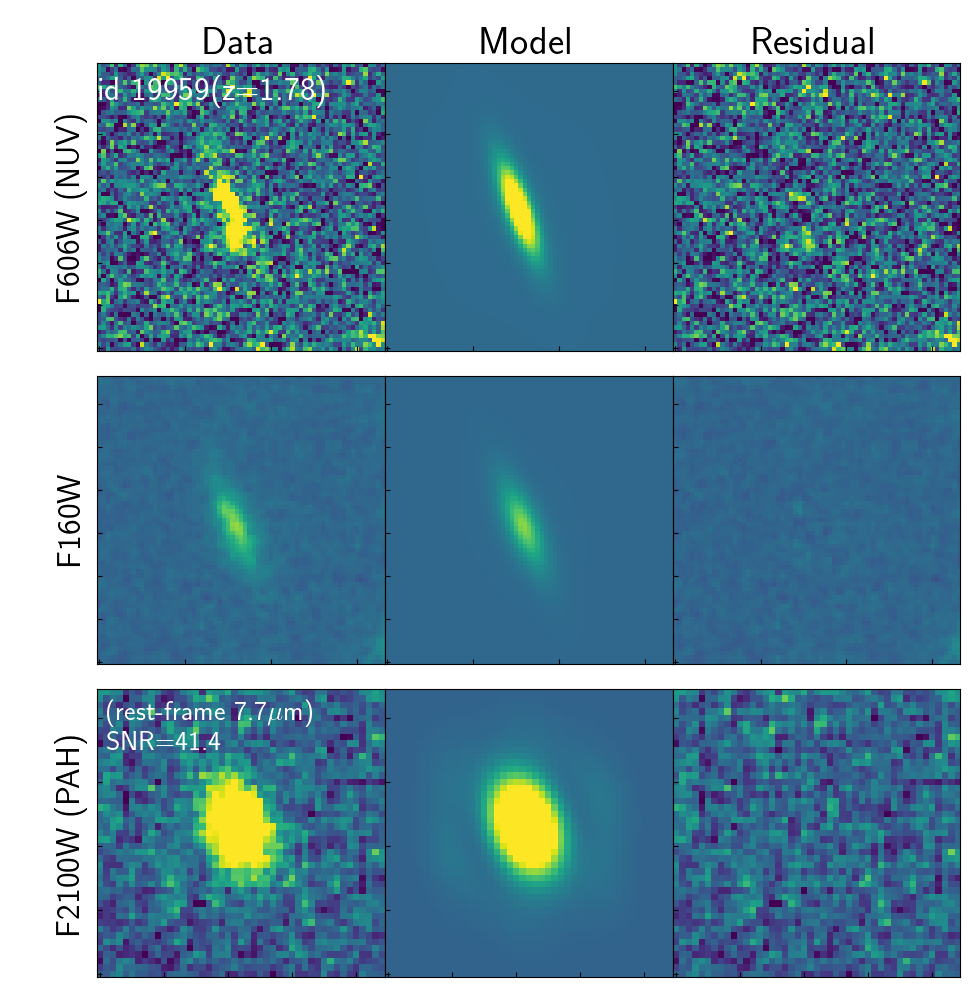}
}
\caption{Examples of the \galfit\ results. For each panel, from \textit{top} to \textit{bottom}, the plots show the \galfit\ measurements in NUV-band, F160W and  PAH-band. For each row, from \textit{left} to \textit{right} are data, \galfit\ model (PSF-convolved) and residual. Each image is $2^{\prime\prime}\times2^{\prime\prime}$ centered on the target galaxies. For each galaxy, the ID and photometric redshift from \citet{Stefanon2017} are labeled. The signal-to-noise ratio of MIRI fluxes are given in their panel. }\label{fig:galfit1}
\end{figure*}

\subsection{The morphology measurements}\label{sec:size}

We use \galfit\ \citep{Peng2002,Peng2010a} for the morphology measurements following the two-\galfit-run approach from \citet{Matharu2019, Matharu2022}, which leads to a high level of agreement with published size measurements (see more details in \citealp{Matharu2019}). 
For each galaxy, we create $10^{\prime\prime}\times10^{\prime\prime}$ cutouts of NUV-, PAH-bands and F160W and associated error images, centered on its F160W coordinates. 
We adopt a single 2D single-component S\'{e}rsic and a background sky profiles to fit each cutout in two iterations. 
A PSF image is included in every \galfit\ fit to account for image resolution limit. 
The error images are used as sigma images when running \galfit. 
In the first iteration, all parameters are kept to free. 
In the second iteration, we fix the shape parameters ($x$,$y$ center, axis ratio, position angle) and sky level to the values from the first iteration, and we re-fit for the effective radius ($R_{\rm eff}$), S\'{e}rsic index ($n$), and magnitude. 
We exclude those \galfit\ outputs marked with an asterisk on either side of any value. 

We visually examine all the \galfit\ results. 
There are seven galaxies that appear to be involved in mergers/interactions, including three galaxies with each having a close companion (id 17353, id 17423, id 20784), four galaxies with evidence of tidal features (id 12363, id 12580, id 17309, and id 20237). 
For the three galaxies with companions, we add an additional single S\'{e}rsic profile for the adjacent galaxy in the \galfit\ input and re-fit with the two-\galfit-run. 
For the four galaxies with evidence of tidal features, their minor components are much fainter than the main structures, thus we use one S\'{e}rsic profile to model the main structure of these galaxies. 
%The median reduced $\chi^2$ of the \galfit\ of the PAH-band, F160W-band and NUV-band are 0.24, 0.51, 0.86. 
Examples of \galfit\ fits are shown in \ref{fig:galfit1}. 
The SNR of the PAH-band (rest-frame 6.2/7.7~$\mu$m) as function of stellar mass are shown in Fig.~\ref{fig:snr}. 90\% of galaxies have successful morphology measurements with SNR of 13 and 12 before and after stellar mass cut at $M_*\geq10^9 M_\odot$. 
Meanwhile, 72\% of galaxies with $M_*\geq10^9 M_\odot$ have successful \galfit\ model fits in PAH-bands. The number of galaxies {with} successful \galfit\ fits in the PAH-, NUV-bands and F160W are listed in Table~\ref{tab:sample}. 

In this paper, we mainly use the $R_{\rm eff}$ (the effective semi-major axis) and $n$ measured from \galfit. 
To account for the covariance of $R_{\rm eff}$ and $n$, we adopt a parameter to describe the fraction of light contained within 1 kpc following \citep{Graham2005a, Graham2005b, Matharu2022, Ji2022} calculated as 
\begin{equation}\label{eq:1}
   f_{\rm 1kpc} = \frac{\gamma(2n, b_{n}R_{\rm eff}^{-1/n})}{\pi},
\end{equation}
where $\gamma(2n, b_{n}R_{\rm eff}^{-1/n})$ is the lower incomplete gamma function and $b_n$ is a $n$-dependent normalization parameter that satisfies $\Gamma(2n) = \gamma(2n, b_{n})$. We adopted approximation values of $b_n$ as functions of $n$ derived in analytical expressions from \citet{Ciotti1999, MacArthur2003}. 

{We employed a Monte-Carlo simulation to estimate the uncertainties of these morphological properties derived from \galfit.   For each iteration we added random noise to each MIRI image using the error image for the original image.  We then rerun \galfit\ by fixing the shape parameters and sky level to the values obtained from the original image, and allowing the $R_{\rm eff}$, $n$ and magnitude to vary. For galaxies without spectroscopic redshift, we also perturb the photometric redshift by drawing a new redshift from a Gaussian distribution with the mean redshift set to $z_{\rm phot}$ and $\sigma$ set to one-half of the difference between the 16th and 84th percentiles of the photometric redshift.   We then re-select the appropriate MIRI band that contains the PAH feature for each galaxy at the ``new'' redshift.  We run the Monte-Carlo with 100 iterations for each band of each galaxy and adopt the 16th and 84th percentiles of the mock $R_{\rm eff}$, $n$ and $f_{\rm 1kpc}$ as their uncertainties on each quantity.  In this way we incorporate the uncertainties of the image, and from the photoemtric redshift into our analysis.}

\subsection{CIGALE SED fitting}\label{sec:cigale}

{We employed the SED fitting Code Investigating GALaxy Emission (\cigale) \citep{Boquien2019, Yang2020} in order to constrain possible AGN contribution to the IR luminosity of our samples,  and to estimate the IR and FUV luminosities in a self-consistent framework that considers the energy balance between the UV/optical and IR. }
{In detail, we adopted a delayed exponential star formation history (SFH) allowing the $\tau$ and stellar age varying from 0.1--10~Gyr and 0.1--10~Gyr, respectively.  
We assumed a \citet{Chabrier2003} IMF and the stellar population synthesis models presented by \citet{Bruzual2003} with solar (Z$_\odot$) metallicity. 
The dust attenuation follows \citet{Calzetti2000}'s extinction law allowing colour excess $E_s$(B-V) to vary from 0 to 0.4. 
The amplitude of the absorption UV bump feature produced by dust at 2175 \AA\ and the slope of the power law are allowed to vary from 0--3 and $-$0.5--0, respectively. 
For the dust emission module, we adopted the dust templates of \citet{Draine2014}. This module models the dust emission with two components, the diffused emission and the photodissociation region (PDR) emission associated with star formation. 
We allow the mass fraction of PAH varying between four different values and the minimum radiation field ($U_{\rm min}$) varying from 0.1, 1.0, 10, 30. }
{For the AGN module, we adopt the SKIRTOR template. 
We retain the default parameters in the AGN module, other than setting the viewing angle $i$ to 30$^{\circ}$ and 70$^{\circ}$ for type 1 and type 2 AGNs, respectively, and a full range of AGN fraction ($f_{\mathrm{AGN}}$) from 0 to 0.9, a fraction that denotes the contribution from the AGN to the total IR luminosity. }
{We adopt the `pdf analysis' method in \textsc{cigale} to compute the likelihood ($\chi^2$) for all the possible combinations of parameters and generate the marginalized probability distribution function (PDF) for each parameter and each galaxy. }

{We run \cigale\ on the photometry measured from the ground-based observations: $u^{*}$, $g^\prime$, $r^\prime$, $i$, $z^\prime$ from Canada–France–Hawaii Telescope (CFHT)/MegaCam and Ks from CFHT/WIRCam, as well as from the space-based observations: six \hst\ bands (F275W, F435W, F606W, F814W, F125W, F160W), three IRAC/Spitzer channels (3.6, 4.5 and 5.8\micron), six \jwst\ MIRI bands (F770W, F1000W, F1280W, F1500W, F1800W and F2100W) and MIPS/Spitzer at 24 $\mu$m and 70 $\mu$m. The ground-based, \hst\ and IRAC data are adopted directly from S17. The MIPS data are included from the catalog provided by \citet{Barro2011}. For galaxies without MIPS detections, we adopted the 5$\sigma$ as an upper limit. Furthermore, we have excluded the IRAC/Spitzer channel at 8 \micron\ is excluded because it is similar to the MIRI F770W, but the latter is substantially deeper. }
 {We adopt the Bayesian results of IR luminosity ($L_{\rm IR}$) and FUV luminosity ($L_{\rm FUV}$) and their associated errors from \cigale\ for estimating IR/UV-based SFR and surface density of SFR (see Section \ref{sec:Sigma}). }
 
{In addition, we adopt the $f_{\rm AGN}$ from \cigale\ to select galaxies that appear to host an AGN with $f_{\rm AGN} \geq 0.1$, and star-formation dominated galaxies with $f_{\rm AGN} < 0.1$ following \citep{Shen2020a}.  \cigale\ calculates $f_{\rm AGN}$ as the fraction of the AGN to the total IR luminosity.   
We found 6 galaxies that likely host AGN based on this criteria. They have a median AGN fraction of 0.20 and AGN fraction in range of 0.17 -- 0.32. We removed these galaxies in our final sample to exclude any possible effect due to AGN contamination. } 

{We observe that the average effective radius in the PAH-band of these AGN candidates is smaller than that of in the F160W (their stellar continuum). The difference between these two is larger than that for our final sample (that excludes AGN candidates). This appears to imply that the presence of an AGN can reduce the measured galaxy sizes, leading to biased conclusions about the morphology of galaxies in the PAH bands. However, because our sample includes only a small number of these AGN candidates, adding them to our sample would not affect the average effective radii of the final sample nor impact our conclusions.  We plan to explore the difference in MIRI morphology of AGN and non-AGN in the future with a larger sample. }

\begin{figure}
\centering
\includegraphics[width=\columnwidth]{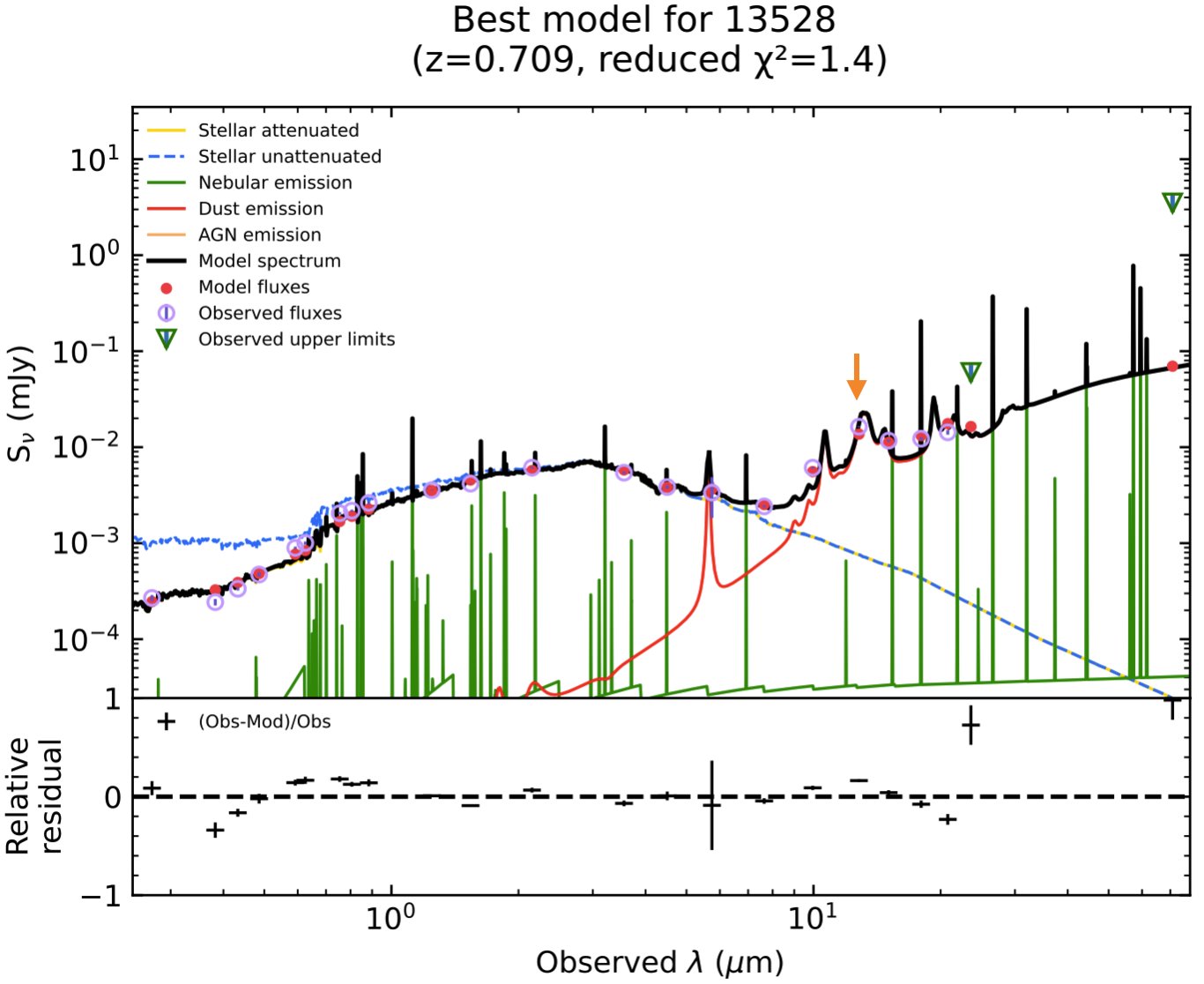}
\includegraphics[width=\columnwidth]{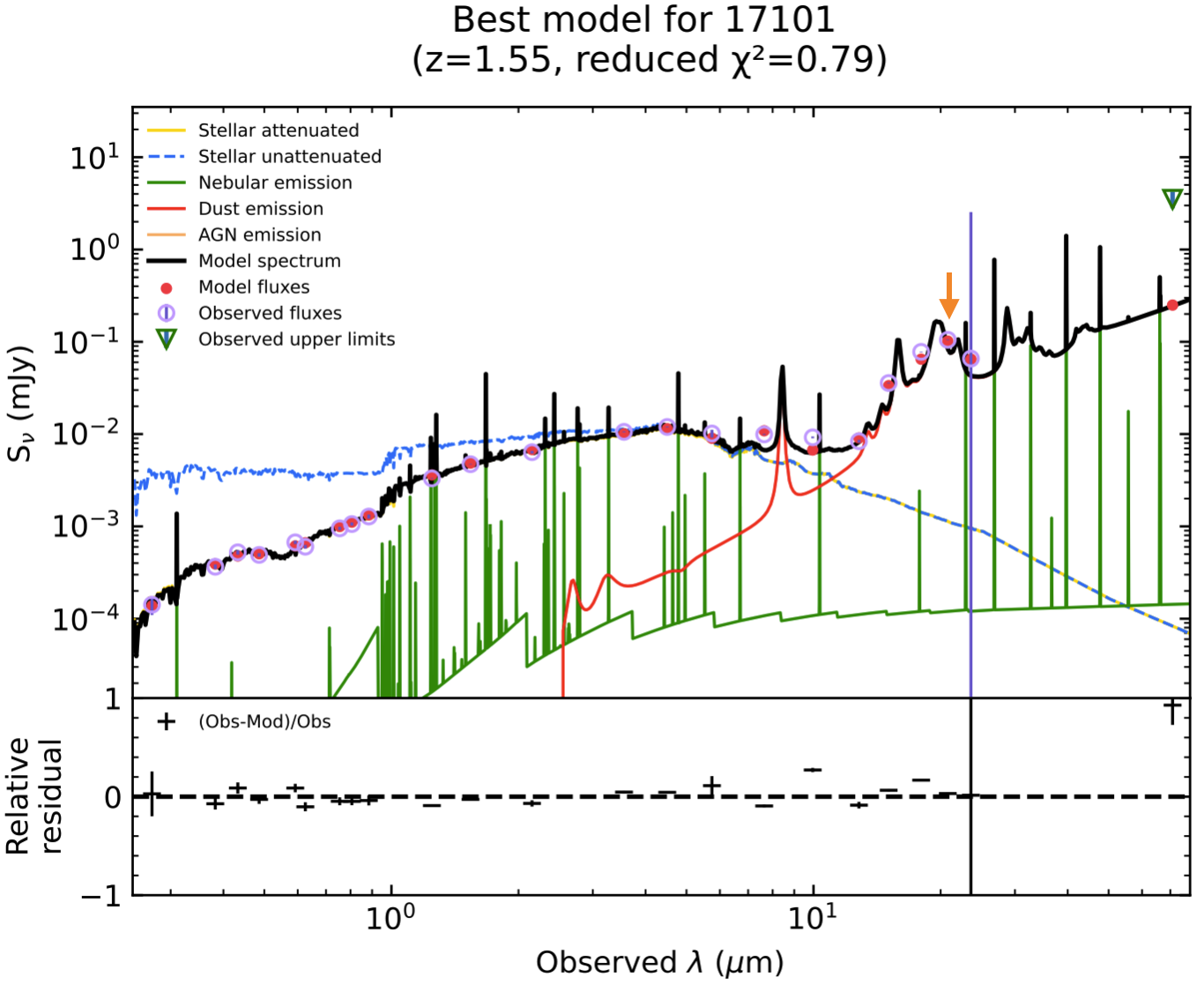}
\caption{{Examples of the best-fitting SED model from CIGALE. In each panel, the top shows the observed photometric fluxes with errors (purple), the 5$\sigma$ upper limit fluxes (green triangles), the CIGALE-derived best model photometry (red dots), and the best-fitting CIGALE model (black). The best-fitting CIGALE model is the sum of the contributions from a dust attenuated stellar emission (yellow; the intrinsic stellar emission is indicated in blue), nebular emission (green), and dust emission (red). The bottom shows the fractional discrepancies between the model and the photometry. The reduced $\chi^2$ of the best-fitting models are indicated in the top labels. The data for the PAH-band are indicated by an orange arrow. In both examples shown here, the PAH-bands contain the rest-frame 7.7\micron\ features. }}\label{fig:cigale}
\end{figure}

\begin{deluxetable}{c|c}
%\tablenum{1}
\tablecaption{{Parameter ranges used in the SED fitting with \textsc{cigale}. }\label{tab:sed}}
\tablewidth{0pt}
\tablehead{
\colhead{Parameter} & \colhead{Values}}
\startdata
\multicolumn{2}{c}{Star formation history (sfhdelayed) }  \\
\hline
$\tau$ [Gyr]  &  0.1, 0.3, 1, 3, 5, 10  \\
Age [Gyr]  & 0.1, 0.5, 1, 3, 5, 7, 10 \\
\hline
\multicolumn{2}{c}{Simple stellar population \citep{Bruzual2003}} \\
\hline
IMF & \citet{Chabrier2003} \\
Metallicity &  0.02 \\ 
\hline
\multicolumn{2}{c}{Dust Attenuation \citep{Calzetti2000}}  \\
\hline
\multirow{2}{*}{E(B-V)$_l$} & 0.01, 0.05, 0.1, 0.2, \\ 
&  0.3, 0.5, 0.7 0.9 \\
E(B-V)$_{\rm factor}$ & 0.44 \\
Amplitude of the UV bump & 0, 1.5, 3 \\
Slope of the power law & -0.5, -0.25, 0 \\ 
\hline
\multicolumn{2}{c}{Dust emission \citep{Draine2014}} \\
\hline
Mass fraction of PAH (\%)& 0.47, 2.50, 4.58, 6.63 \\
Minimum radiation field & 0.1, 1, 10, 30 \\
Power slope $dU/dM \propto U^{-\alpha}$ & 2.0 \\
Dust fraction in PDRs & 0.05 \\
\hline
\multicolumn{2}{c}{AGN emission (SKIRTOR)} \\
\hline
Viewing angle & 30, 70 \\
$f_{\mathrm{AGN}}$ & 0 - 0.9 \\
\enddata
%\tablecomments{}
\end{deluxetable}

\subsection{Surface density of the Stellar Mass and the SFR} \label{sec:Sigma}

We further use the morphology parameters measured from \galfit\ to derive surface densities of stellar mass and SFR in our analyses to support the comparison between unobscured SFR, obscured SFR and stellar mass.  
The surface density of stellar mass within effective radius ($\Sigma_{\rm eff}$) and within 1 kpc ($\Sigma_{\rm 1kpc}$)  are calculated following {\citep{Cheung2012, Barro2017a, Matharu2022}} as
\begin{equation}
    \Sigma_{\rm eff, M_*} = \frac{0.5 \rm M_*}{\pi R^2_{\rm eff}},
\end{equation}
\begin{equation}
    \Sigma_{\rm 1kpc, M_*} = M_*\times f_{\rm 1kpc}
\end{equation}
where $M_*$ is the stellar mass, $R_{\rm eff}$ is the effective radius, and $f_{\rm 1kpc}$ is the fraction of light contained within 1 kpc defined in eq.~\ref{eq:1} both measured in F160W.  An advantage of using $\Sigma_{\rm 1kpc}$ is that it is more robust against the covariance between the effective size and S\'{e}rsic index, because the $\Sigma_{\rm 1kpc}$ is the integral over the surface brightness (see, e.g., the discussion in \citealt{Estrada-Carpenter2020,Matharu2022}).

We convert {the IR and FUV luminosities from \cigale} to SFR using calibrations from the literature for the UV SFR (SFR$_{\rm UV}$) and IR SFR (SFR$_{\rm IR}$) following \citet{Kennicutt2012}:
\begin{equation}
    \rm SFR_{IR} =  \cal{K}_{\rm IR} \times L_{\rm IR},
\end{equation}
\begin{equation}
    \rm SFR_{UV} = \cal{K}_{\rm UV} \times L_{\rm FUV},
\end{equation}
where we adopt ${\cal K}_{\rm IR} = 10^{-43.41} \rm M_\odot\ yr^{-1} ergs^{-1}\ s$ and ${\cal K}_{\rm UV} = 10^{-43.35} \rm M_\odot\ yr^{-1} ergs^{-1}\ s$ \citep{Madau2014}. As noted in \citet{Kennicutt2012}, these constants are appropriate for a Chabrier IMF.

We then calculate the $\Sigma_{\rm eff, SFR}$ and $\Sigma_{\rm 1kpc, SFR}$ as 
\begin{equation}
    \Sigma_{\rm eff, SFR} = \frac{0.5\rm \times SFR}{\pi R^2_{\rm eff}},
\end{equation}
\begin{equation}
    \Sigma_{\rm 1kpc, SFR} = \rm SFR\times f_{\rm 1kpc},
\end{equation}
where we use the UV-based SFR with the $R_{\rm eff}$ and $f_{\rm 1kpc}$ measured in the NUV-band, and the IR-based SFR with the $R_{\rm eff}$ and $f_{\rm 1kpc}$ measured in the PAH-band to obtain their respective surface densities. {We note that the UV-based SFR is uncorrected for dust attenuation, and therefore it represents the dust-unobscured portion of the galaxies' star formation. }

\section{Results} \label{sec:result}

In this section, we first explore differences among morphologies measured in NUV-, PAH-bands and F160W. 
We focus on the effective radius, S\'{e}rsic index, and the fraction of light contained within 1 kpc in order to compare the spatial extent, the shape of light profile, and the concentration of different tracers of the light (of the unobscured star-forming regions, stellar continuum, and obscured star-forming regions, respectively).
We compare the effective radius and S\'{e}rsic index of these three bands for MIRI-detected SFGs in Section~\ref{sec:reff} and \ref{sec:n}, respectively. 
In Section~\ref{sec:mass-morph}, we present the differences of effective radius, S\'{e}rsic index and fraction of light contained within 1 kpc between PAH-, NUV-bands and F160W as a function of stellar mass.

Secondly, we explore the surface density of obscured and unobscured SFR (within effective radius and within 1 kpc) and the obscured fraction of star formation in the center of galaxies. 
In Section \ref{sec:mass-Sigma}, we compare the surface density of obscured-SFR derived from $\rm SFR_{IR}$ and PAH-band morphology with the surface density of unobscured-SFR derived from $\rm SFR_{UV}$ and NUV-band morphology and the surface density of stellar mass derived from $M_*$ and F160W morphology. 
The comparison between the obscured fractions of star formation within 1 kpc and integrated over the entire galaxy are further present in Section \ref{sec:fobs}.

\subsection{On the Effective Radii of MIRI-detected SFGs}\label{sec:reff}

\begin{figure*}
    \centering
    \includegraphics[width=\textwidth]{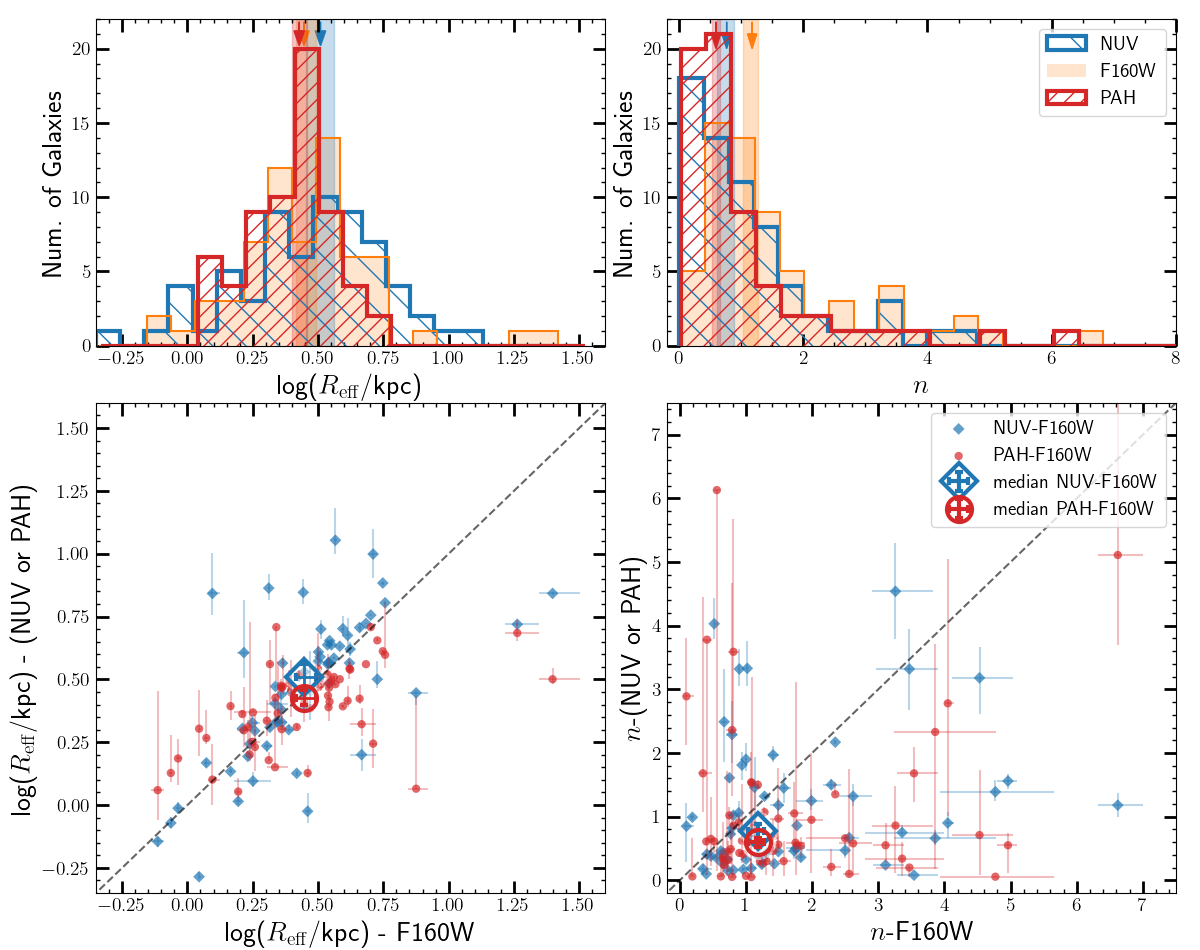}
    \caption{\textit{Top:} The effective radius ($R_{\rm eff}$, \textit{left}) and S\'{e}rsic index ($n$, \textit{right}) histograms of NUV-band (blue), F160W ({orange}) and PAH-band ({red}) for the final sample of MIRI-detected SFGs.  Median values and associated uncertainties of these three bands are marked by arrows and shaded regions in the same color as their histograms. \textit{Bottom:} The $R_{\rm eff}$ (\textit{left}) and $n$ (\textit{right}) of NUV-band (blue diamonds) and PAH-band ({red} dots) versus F160W. Median values and uncertainties are marked by large open markers with errorbars in the same color as data points. }
    \label{fig:size_n}
\end{figure*}

The left panels of Fig.~\ref{fig:size_n} show the effective radius histograms of NUV- and PAH-bands and F160W and scatter plots of NUV-/PAH-band versus F160W for the final sample of MIRI-detected SFGs. 
{To account for uncertainties on each measured morphological properties, we adopt a bootstrap method.  For each bootstrap iteration, we randomly draw, with replacement, the same number of galaxies from the final sample, and for each galaxy, we randomly sample a mock value from a Gaussian with the measured value as the mean and the one-half of the difference between the 16th and 84th percentiles of as the standard deviation. We then obtain the median from the distribution of the bootstraped values. We repeat this process for 1000 iterations. We adopt the median and the 16th/84th percentiles of these bootstrap median values as final median and associated error. }
The medians of the effective radius and {errors} on the median are marked by arrows and shaded regions in the top panel and by large open markers with errorbars in the bottom panel. 

The median $R_{\rm eff}$ of the NUV-band, F160W and PAH-band are {$3.2\pm0.4$ kpc, $2.8\pm0.3$ kpc, and $2.7\pm0.2$ kpc,} respectively (also see Tab.~\ref{tab:props}). 
{It appears that the PAH-band sizes are, on average, similar to the F160W sizes, but they are both smaller than the NUV-band sizes. }

We employed the Kolmogorov–Smirnov statistic (KS) test and Mann-Whitney U (MWU) test and their resultant p-values to determine a likelihood that the the NUV-, PAH-bands and F160 size distributions are consistent with the same parent distribution. We adopt a p-value = 0.05 as the significance threshold\footnote{If p-value $<$ 0.05, the probability of the two distributions drawing from the same distribution is very small. Otherwise, we cannot reject the null hypothesis that the two distributions are drawn from the same distribution.}. 
Both the KS and MWU statistic calculate a probability that two distributions are drawn from the same distribution. The MWU test is more sensitive to differences in medians, while the KS test is more sensitive to differences in the cumulative distributions of the two samples. 
The p-values of KS and MWU tests on the $R_{\rm eff}$ distributions between each two bands are summarized in Tab.~\ref{tab:tests}. The p-values on the same pair of bands between these two tests are mostly consistent. 

{Both tests return $p$-values $\le0.05$ between the $R_{\rm eff}$ distributions of the PAH- and NUV-bands. We therefore reject the null hypothesis that the spatial extent of obscured- and unobscured-star formation are drawn from the same parent distribution. }
{However, The KS and MWU test between the F160W sizes and the star-formation ones are inconclusive: based on these tests, there is no significant difference between the $R_\mathrm{eff}$ distributions of the NUV and stellar continuum bands, nor between the PAH and stellar continuum bands.  }
Therefore, we reject the null hypothesis that the NUV- and PAH-band sizes come from the same distribution, but we can not do rule out that this is the case for the PAH-/NUV-band and F160W sizes.

\begin{deluxetable}{ccccc}
\tablecolumns{5}
\tablecaption{Properties of samples \label{tab:props} }
\tablehead{ \colhead{Band} & \colhead{$\langle z \rangle$} & \colhead{$\langle \rm log(M_*/M_\odot) \rangle$} & \colhead{$\langle R_{\rm eff}/kpc \rangle$} & \colhead{$\langle n \rangle$} \\ [-4pt]
\colhead{(1)} & \colhead{(2)} & \colhead{(3)} & \colhead{(4)} & \colhead{(5)} }
\startdata
PAH & \multirow{3}{*}{$1.2_{-0.5}^{+0.6} $} & \multirow{3}{*}{{$9.7_{-0.5}^{+0.3}$}} & $2.7{{\pm 0.2}}$ & $0.6{{\pm 0.1}}$ \\
F160W &  &  & $2.8{{\pm 0.2}}$ & {$1.1 \pm 0.1$} \\
NUV &  &  & $3.2{{\pm 0.4}}$ & $0.8{{\pm 0.1}}$ \\
\enddata
\tablecomments{ (1) The name of band; (2) and (3) The median and 16th/84th percentiles of redshift and stellar mass of the final sample; (4) and (5) The median and {uncertainties on median} of effective radius and S\'{e}rsic index measured in each band {using the bootstrap method}. }
\end{deluxetable}

\begin{deluxetable}{c|ccccc}
\tablecolumns{5}
\tablecaption{P-values from statistic tests\label{tab:tests}}
\tablehead{\colhead{Comparison} & \multicolumn{2}{c}{$R_{\rm eff}$} & \colhead{} & \multicolumn{2}{c}{$n$} \\[-2pt] 
\cline{2-3} \cline{5-6}
\colhead{Bands} & \colhead{KS} & \colhead{MWU} & \colhead{} & \colhead{KS} & \colhead{MWU} \\ [-4pt]
\colhead{(1)} & \colhead{(2)} & \colhead{(3)} & \colhead{} & \colhead{(4)} & \colhead{(5)} } 
\startdata
PAH vs.\ F160W & {0.09 (0.22)} & {0.22} & & {$10^{-5}$ (0.44)} & {$10^{-5}$}\\
PAH vs.\ NUV & {$10^{-3}$ (0.36)} & {0.03} & & {0.42 (0.16)} & {0.35}\\
F160W vs.\ NUV & {0.21 (0.19)} & {0.38} & & {$10^{-3}$ (0.31)} & {$10^{-3}$}\\
\enddata
\tablecomments{(1) The name of two tested bands; (2) and (4) The p-values from KS tests; {The KS statistic are shown in the parentheses};  (3) and (5) The p-values from Mann-Whitney U (MWU) tests. {We adopt a p-value $<$ 0.05 as rejecting the hypothesis that the two distributions drawn from the same distribution}.}
\end{deluxetable}

\subsection{On the S\'{e}rsic indexes of MIRI-detected SFGs}\label{sec:n}

In right panels of Fig.~\ref{fig:size_n}, we show the S\'{e}rsic index histograms of NUV-, PAH-bands and F160W, and scatter plots of NUV-/PAH-band versus F160W. 
The S\'{e}rsic indexes of F160W are, on average, larger than those of NUV-band and PAH-bands. 
The median S\'{e}rsic indexes of NUV-band, F160W and PAH-band are {$0.8\pm0.1$, $1.1\pm0.1$, and $0.6\pm0.1$}, respectively (also listed in Tab.~\ref{tab:props}). 
These median values suggest that the stellar continuum prefer a disk-like morphology ($n\sim1$; and this is explored further by Magnelli et al. \textit{in prep.}).  
In contrast, the obscured-star formation (PAH-band) and unobscured-star formation (NUV-band) profiles prefer a surface-brightness profile that is flatter within the effective radius (this is implied by the lower S\'ersic indexes).  

In addition, the KS and MNU tests reveal that the S\'{e}rsic index distribution of F160W is significantly different from those of PAH- and NUV-bands, while the S\'ersic index distributions of NUV- and PAH-bands are likely drawn from the same parent distribution. 
We interpret this as evidence that the obscured- and unobscured-star formation follow a profile with similar shape which is different from the profile of stellar continuum. We discuss this further below.

\subsection{On the Mass-Morphology relation of MIRI-detected SFGs}\label{sec:mass-morph}

\begin{figure*}
    \centering
    \includegraphics[width=\textwidth]{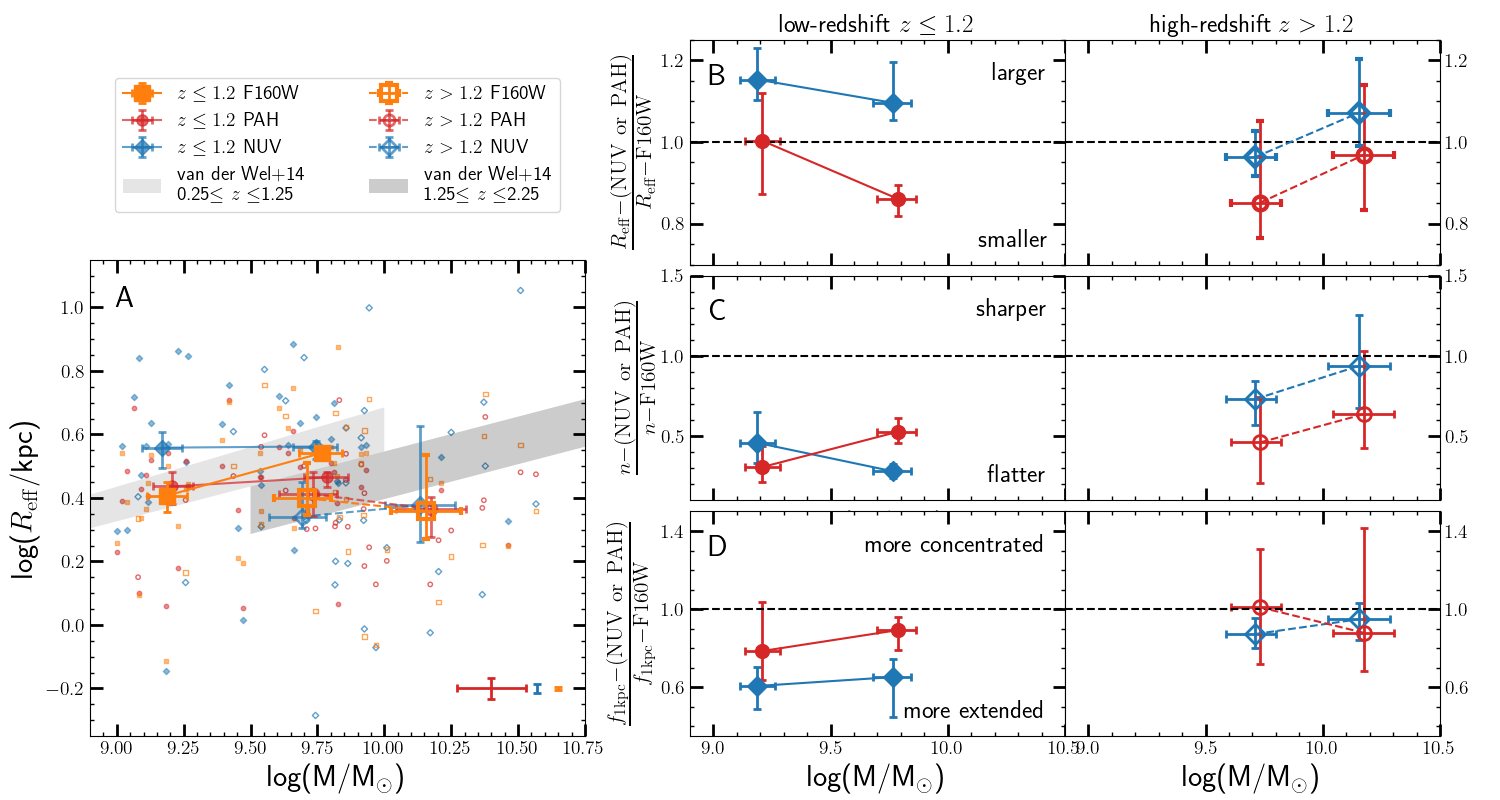}
    \caption{Morphological measurements versus stellar mass for the final sample of MIRI-detected SFGs. \textit{Left:} the effective radius-mass relation with size measured in the NUV-band (blue), F160W (green) and PAH-band (orange). The larger colored data points with error bars show the median values and associated uncertainties {obtained from the bootstrap method (Section \ref{sec:reff})}. The solid makers denote the low-redshift ($z\leq1.2$) subsample and open markers denote the high-redshift ($z>1.2$) subsample. {The median of $R_{\rm eff}$ uncertainties of the three bands are indicated by colored errorbars, as well as the median of stellar mass uncertainty. } \textit{Right:} the top, middle, and bottom panels show the ratio of the effective radius, S\'{e}rsic index and $f_{\rm 1kpc}$ measured in the NUV- or PAH-band to the same quantity in the F160W band in bins of stellar mass{, separated in to low-redshift and high-redshift subsamples. } We shifted median values slightly along the $x$-axis in {each} panel for clarity. }
    \label{fig:size_mass_z}
\end{figure*}

Previous studies have measured size evolution for SFGs in their effective radii (i.e., in the stellar continuum band $R_{\rm eff} \propto (1+z)^{-0.75}$, see \citealt{vanderWel2014}).  We therefore separate our MIRI-detected SFGs into a low-redshift subsample ($z\leq1.2$) and a high-redshift subsample ($z>1.2$). 
In Fig~\ref{fig:size_mass_z}, we plot the effective radii of galaxies measured from their NUV-, PAH-bands and F160W as a function of their stellar mass (as blue, {red and orange} data points for the three bands, as labeled, with solid makers denoting the low-redshift subsample and open markers denoting the high-redshift subsample). 
Each subsample is then binned by stellar mass to show the average difference between these bands. 
The low-/high-redshift subsample selection and binning are chosen to have a similar number of galaxies ($\sim$ {15}) in each bin. 
The median effective radius {and stellar mass} and associated errors {obtained from the bootstrap method (see Section \ref{sec:reff})} are shown as larger markers with error bars in the same color convention as data points for individual measurements. 
The median properties of galaxies in each subsample and mass bin are listed in Tab.\ref{tab:bins}. 

The stellar size-mass relation for late-type galaxies at a rest-frame wavelength of $\sim$5000\AA\ from \citet{vanderWel2014} is shown as the light and dark grey shaded regions in redshift ranges of $z=$0.25-1.25 and $z=$1.25-2.25, respectively. 
The median sizes derived from the F160W band are consistent with the size-mass relations from \citet{vanderWel2014}  for both low-redshift and high-redshift subsamples. 
{We notice that our median F160W size of the high-mass bin for the high-redshift subsample is slightly offset from the size-mass relations, though it is consistent within its uncertainty. This could suggest the galaxies sample is slightly biased in this mass bin toward galaxies with smaller sizes measured in their stellar continuum, but this would need to be confirmed with larger samples. } 
{We also warn the reader that this size -- mass relation (and following morphology -- mass relations) are based on a single redshift/mass binning that does not account for uncertainties on stellar mass, redshift and sample incompleteness.  We will discuss this further in Section \ref{sec:dis_caveat} }

However, we see that the $R_{\rm eff}$--mass relation for the PAH- and NUV-bands show different evolution compared to size--mass relations for the F160W and from \citet{vanderWel2014}{, especially for the low-redshift subsample}. 
We further investigate the morphological difference of PAH- and NUV-bands relative to F160W as a function of stellar mass, in terms of effective radius in Section \ref{sec:size-mass}, S\'{e}rsic index in Section \ref{sec:n-mass} and fraction of light contained within 1 kpc in Section \ref{sec:f-mass}.

\begin{deluxetable*}{ccccccccccccc}
\tablecolumns{13}
\tablecaption{Median properties of galaxies in each subsample/bin as shown in Fig.~\ref{fig:size_mass_z} \label{tab:bins} }
\tablehead{\colhead{Subsample} & \colhead{Mass Bin} & \colhead{Num.}& \colhead{$\langle z \rangle$} & \colhead{$\langle \rm{log(M_*/M_\odot)}\rangle$} & \multicolumn{2}{c}{PAH-band}& \colhead{} & \multicolumn{2}{c}{NUV-band} & \colhead{} & \multicolumn{2}{c}{F160W} \\[-2pt]
\cline{6-7} \cline{9-10} \cline{12-13} 
\colhead{} & \colhead{} & \colhead{of} & \colhead{} & \colhead{} & \colhead{$\langle R_{\rm eff} \rangle$} & \colhead{$\langle n\rangle$} & \colhead{} & \colhead{$\langle R_{\rm eff} \rangle$} & \colhead{$\langle n\rangle$} & \colhead{} & \colhead{$\langle R_{\rm eff} \rangle$} & \colhead{$\langle n\rangle$} \\[-4pt]
\colhead{} & \colhead{$\rm \log(M_\ast/M_\odot)$} & \colhead{Gals}& \colhead{} & \colhead{} & \colhead{kpc} & \colhead{} & \colhead{} & \colhead{kpc} & \colhead{} &\colhead{} & \colhead{kpc} & \colhead{}\\[-4pt]
\colhead{(1)} & \colhead{(2)} & \colhead{(3)} & \colhead{(4)} &\colhead{(5)} & \colhead{(6)} & \colhead{(7)} & \colhead{} & \colhead{(8)} & \colhead{(9)} & \colhead{} & \colhead{(10)} & \colhead{(11)}}
\startdata
\multirow{2}{*}{$z\leq1.2$} & {$<9.5$} & {18} & {0.7$\pm$0.1} & {9.2$\pm$0.1} & {2.7$\pm$0.3} & {0.5$\pm$0.2} & & {3.6$\pm$0.5} & {0.6$\pm$0.2} & & {2.6$\pm$0.3} & 1.1$\pm$0.1\\
& {$\geq9.5$} & {16} & {0.7$\pm$0.1} & {9.8$\pm$0.1} & {2.9$\pm$0.2} & {0.6$\pm$0.1} & & {3.6$\pm$0.2} & {0.4$\pm$0.2} & & {3.5$\pm$0.1} & 1.3$\pm$0.2\\
\hline
\multirow{2}{*}{$z>1.2$} & {$<9.9$} & {15} & {1.4$\pm$0.5} & {9.7$\pm$0.1} & {2.6$\pm$0.4} & {0.7$\pm$0.4} & & {2.2$\pm$0.4} & {0.8$\pm$0.1} & & {2.5$\pm$0.5} & {1.4$\pm$0.8}\\
& {$\geq9.9$} & {15} & {2.0$\pm$0.6} & {10.2$\pm$0.1} & {2.3$\pm$0.3} & 0.6$\pm$0.2  & & {2.4$\pm$1.0} & {1.2$\pm$0.3} & & {2.3$\pm$0.7} & {1.0$\pm$0.2} \\
\enddata
\tablecomments{(1) The redshift cut of each subsample; (2) The stellar mass range of mass bin; (3) The number of galaxies in each mass bin; (4) and (5) The median and associated uncertainties of redshift and stellar mass for each bin; (6) - (11) The median and associated errors of $R_{\rm eff}$ and $n$ in PAH-, NUV-band and F160W for each bin. {The median values are obtain from the bootstrap method that also account for uncertainties on morphologies. The errors are the one-half of the difference between the 16th and 84th percentiles of the bootstrap median values. } }
\end{deluxetable*}

\subsubsection{The Size---Mass Relations}\label{sec:size-mass}

Fig.~\ref{fig:size_mass_z} {panel B} compares the median ratio of sizes measured in the PAH- and NUV-bands to those in the F160W in bins of stellar mass. 
The sizes of obscured and unobscured star-formation show different relations to the size of the stellar continuum (measured by the F160W band). Specifically, at all masses, the median PAH/F160W size ratio is {similar} to or \textit{smaller} than unity, while the median NUV/F160W size ratio is {similar} to or \textit{larger} than unity. 

{For the low-redshift subsample, the median PAH-band size is similar to the F160W size in the low-mass bin, but the median PAH-band size is smaller than that of F160W in the high-mass bin with a 4$\sigma$ significant level.   For the high-redshift subsample, the median PAH-band sizes are not significantly different from the F160W sizes in either of the stellar mass bins:  the differences between the sizes in the difference bands is within their uncertainties in both mass bins.}

{We also observed that at fixed \textit{stellar mass} of $\log (\rm M_*/M_\odot) = 9.5 - 10$ across the two redshift subsamples (i.e., the high-mass bin of the low-redshift subsample and the low-mass bin of the high-redshift subsample), the ratio of the sizes in the PAH-band to F160W are consistent to be $\sim$0.9 (see Figure~\ref{fig:size_mass_z}). For galaxies at lower stellar mass ($\log (\rm M_*/M_\odot) = 9 - 9.5$), this ratio is unity. Despite the large errors on these ratios, these results tentatively suggest that as galaxies increase their stellar mass, the extent of the obscured star formation, traced by the PAH band, is decreasing. We discuss this further in the Section~\ref{sec:dis_SF}.}

{On the other hand, }the NUV-band sizes are larger than F160W sizes for both stellar mass bins for the low-redshift subsample. This is significant at {2$\sigma$} for the low-mass bin. The NUV sizes for the high-redshift subsample are similar to the F160W sizes (within 1$\sigma$ uncertainties). 

Assuming that dust attenuation is responsible for some of the differences, and if dust is concentrated at the center of galaxies, the observed effective radius of galaxies in the NUV-bands and stellar continuum should be larger than their intrinsic effective radius (e.g., \citealt{Nelson2016a}). 
This could account for the modestly larger NUV sizes we find in the two higher mass bins of both redshift subsamples \citep{Suess2019, Suess2021}. 
However, the obscured fraction of star formation is relatively small in low-mass galaxies as compared to massive galaxies, indicating the amount of dust is lower in these low-mass galaxies (see more discussion in Section~\ref{sec:fobs}). Thus, the effect of dust cannot solely explain the much larger unobscured star formation at the lower mass bin of the low-redshift subsample. 
We further discuss this in Section~\ref{sec:mass-Sigma}. 
Thus, for these low-mass galaxies, their star formation might be intrinsically larger than stellar continuum. % and obscured SF regions. 

\subsubsection{The S\'{e}rsic index---Mass Relations}\label{sec:n-mass}

The S\'{e}rsic index can inform the shape of profiles as a larger $n$ corresponding to a sharper profile within the effective radius, while a smaller $n$ corresponding to a flatter profile within the effective radius. 
We compare the median ratios of the S\'{e}rsic indexes measured in the PAH- and NUV-bands to those in the F160W in bins of stellar mass in {the panel C of} Fig.~\ref{fig:size_mass_z}.  
The S\'ersic indexes measured in the PAH-band are consistently smaller than those of F160W across the stellar mass range and for both high-redshift and low-redshift subsamples. This indicates that the surface brightness profiles of obscured-star formation are generally flatter than those of stellar continuum (at least within the effective radius). As illustrated in Fig.~\ref{fig:size_n}, the fact that the average S\'ersic index of galaxies in the PAH-bands are $n \simeq 0.6$, which indicates that the PAH-emission follows profiles that are flatter than those of stellar-continuum with an average $n \simeq 1.2$ which may trace both disk and bulge components.  

Similarly, the median S\'{e}rsic indexes measured in the NUV-band for the low-redshift subsample are smaller than those of F160W and largely consistent with those measured in the PAH band. This result suggests that for galaxies at $z\leq1.2$, the profiles obscured and unobscured star-formation in these galaxies follow a similar shape.  However, the median ratios of the S\'{e}rsic index measured in the NUV-band to those in the F160W for the high-redshift subsample are closer to unity, which suggests there might be a redshift evolution in the morphological profiles in these different wavelengths or other bias (see next section). 

\subsubsection{The $f_{\rm 1kpc}$---Mass Relations}\label{sec:f-mass}

Finally, we compare the fraction of light contained within 1 kpc measured in the PAH- and NUV-bands relative to that of F160W ($f_\mathrm{1~kpc}$ using Eq.~\ref{eq:1}). 
As shown in {the panel D of} Fig~\ref{fig:size_mass_z}, the median ratios of $f_{\rm 1kpc}$ measured in the PAH-band to the F160W band is close to unity (within their uncertainties).  This is true at all redshifts and stellar masses.  This suggests the concentration of obscured-star formation and the stellar light are similar. %

However, the median $f_{\rm 1kpc}$ measured in the NUV-band is significantly smaller than those measured in the F160W band for the low-redshift subsample {at 3$\sigma$ and 2$\sigma$ significant levels}, suggesting the unobscured-star formation is more extended than stellar continuum. The median ratios of $f_{\rm 1~kpc}$ between the NUV-band to F160W is closer to unity for the high-redshift subsample, implying there is either real redshift evolution or some bias in our choice of bandpasses at the different redshifts.   

Combining the differences of NUV-band morphologies (i.e., $R_{\rm eff}$, $n$ and $f_{\rm 1kpc}$) between the low-redshift and high-redshift subsamples, we suspect that our NUV morphology measurements are dominated by the clumps for these high-redshift galaxies or massive galaxies, possibly due to their high dust attenuation, rather than the full extend of unobscured-star formation for those low-redshift galaxies or low-mass galaxies. 
However, due to the small sample size and the incomplete sample in the mass -- redshift space, we cannot draw a conclusion on whether this effect is a mass-dependence or due to evolution. 
However, this could also due to some bias in our choice of bandpasses at the different redshifts. 
{At $z\sim 1.1$, the NUV-band is referring to rest-frame wavelengths of 0.21 \micron, while the stellar continuum is referring to 0.76 \micron. However at $z\sim 2.5$, the NUV-band is referring to rest-frame wavelengths of 0.17 \micron, while the stellar continuum is referring to 0.45 \micron. This could in part explain the stellar continuum morphology appears to be similar to the NUV-band morphology in the high-mass, high-redshift bin.}

\subsection{The Surface Density---Mass Relations} \label{sec:mass-Sigma} 

\begin{figure*}
    \centering
    \includegraphics[width=\textwidth]{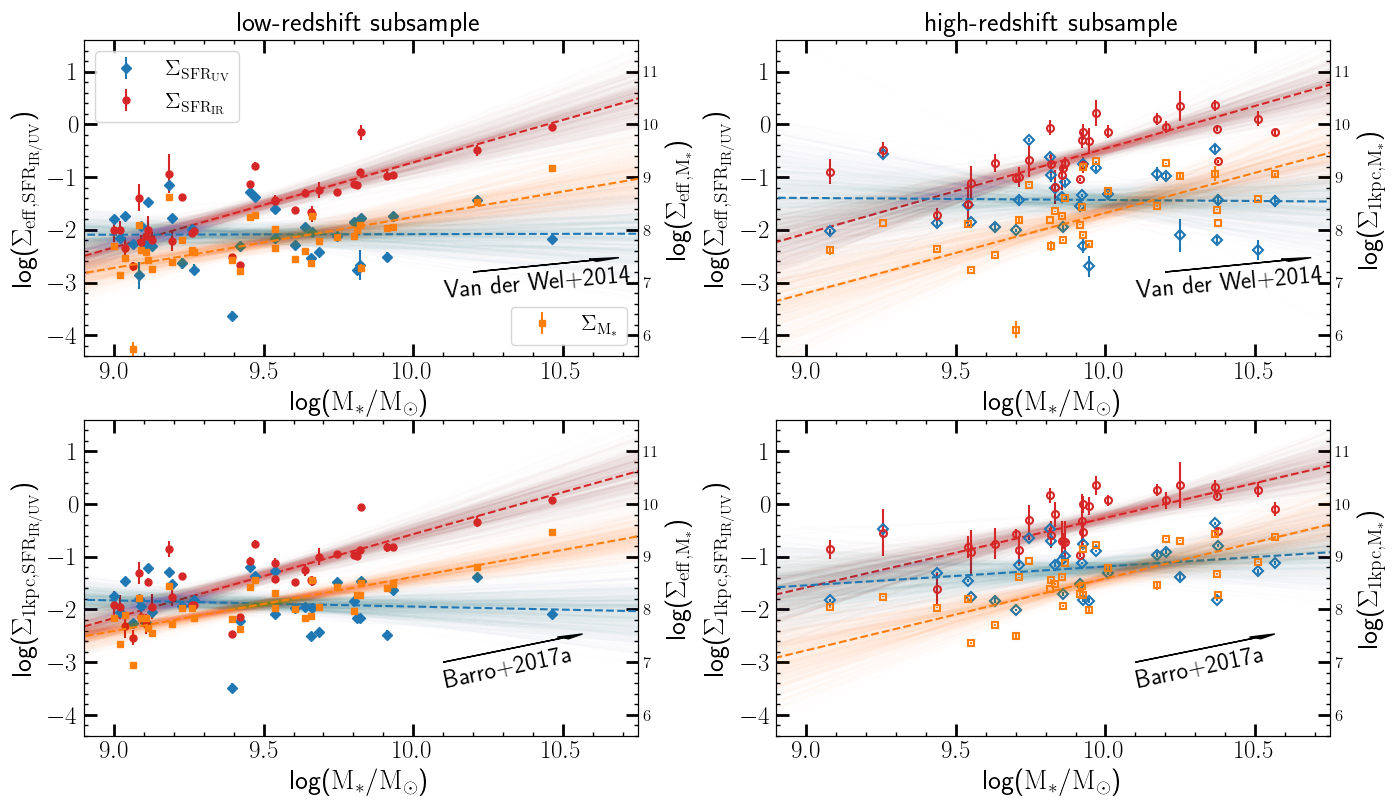}

    \caption{Surface density within the effective radius ($\Sigma_{\rm eff}$, \textit{top}) and within 1 kpc ($\Sigma_{\rm 1kpc}$, \textit{bottom}) of SFR$_{\rm IR}$, SFR$_{\rm UV}$, and stellar mass as functions of stellar mass, and separated into low-redshift (\textit{left}) and high-redshift (\textit{right}) subsamples. 
    In each panel, the small colored data points are the measurements for individual galaxies.  {The dashed color lines are best-fitted linear relations, along with 400 randoms fits from posterier. } 
    The black arrows are the mean growth from the literature \citep{vanderWel2014, Barro2017a}. Note that the zeropoints of the ordinates for surface density of SFR and surface density of stellar mass are different. However, these ordinates span the same range for both surface densities ($\Sigma_{\rm SFR}$ and $\Sigma_{\rm M_*}$). Therefore, the slope of these relations can be directly compared but not the absolute values. }
    \label{fig:sigma}
\end{figure*}

\begin{deluxetable*}{cccccccccc}
\tablecolumns{10}
\tablecaption{The fits of the surface density of stellar mass, IR-based and UV-based SFRs -- mass relations in form of $\rm log \Sigma = \alpha (log(M_*/M_\odot) - 9) + log(A)$ \label{tab:slopes}}
\tablehead{\colhead{$\Sigma$-mass} & \colhead{subsample} & \multicolumn{2}{c}{$\Sigma_{\rm SFR_{IR}}$} & \colhead{} & \multicolumn{2}{c}{$\Sigma_{\rm SFR_{UV}}$} & \colhead{} & \multicolumn{2}{c}{$\Sigma_{*}$} \\ [-2pt]
\cline{3-4} \cline{6-7} \cline{9-10} 
\colhead{} & \colhead{} & \colhead{$\alpha$} & \colhead{log(A)} & \colhead{} &\colhead{$\alpha$} & \colhead{log(A)} & \colhead{} & \colhead{$\alpha$} & \colhead{log(A)}\\ [-4pt]
\colhead{(1)} & \colhead{(2)} & \colhead{(3)} & \colhead{(4)} & \colhead{} &\colhead{(5)} & \colhead{(6)} & \colhead{} & \colhead{(7)} & \colhead{(8)}} 
\startdata
\multirow{2}{*}{$\Sigma_{\rm eff}$} & low-$z$ & {1.6$\pm$0.3} &  {-2.3$\pm$0.2} & &  {0.0$\pm$0.3} &  {-2.1$\pm$0.2} & &  {1.0$\pm$0.3} &  {7.3$\pm$0.2}\\
& high-$z$ &  {1.6$\pm$0.3} &  {-2.1$\pm$0.3} & &  {-0.0$\pm$0.5} &  {-1.4$\pm$0.5} & &  {1.5$\pm$0.5} &  {6.8$\pm$0.5} \\
\hline
\multirow{2}{*}{$\Sigma_{\rm 1kpc}$} & low-$z$ &  {1.5$\pm$0.5} &  {-2.2$\pm$0.2} & &  {-0.1$\pm$0.3} &  {-1.8$\pm$0.2}  & &  {1.0$\pm$0.2} & 7.6$\pm$0.1\\
& high-$z$ &  {1.3$\pm$0.3} &  {-1.6$\pm$0.3} & &  {0.4$\pm$0.4} &  {-1.5$\pm$0.4} & &  {1.4$\pm$0.4} &  {7.2$\pm$0.4}
\enddata
\tablecomments{(1) The name of $\Sigma$ within effective radius or within 1 kpc; (2) The name of subsample; (3) - (8) The slopes and intercepts of the fitted $\Sigma$-mass relations {using linmix \citep{Kelly2007}. The best-fitted parameters are the median of 400 fitted parameters of random draws from the posterior, and the associated errors are one-half of the difference between the 16th and 84th percentiles of the fitted parameters.} }
\end{deluxetable*}

In this section, we compare the surface density of SFR derived from IR and UV and the surface density of stellar mass, as shown in Fig.~\ref{fig:sigma}. 
To better quantify the relations, we plot the $\Sigma_{\rm SFR_{IR}}$, $\Sigma_{\rm SFR_{UV}}$ and $\Sigma_{\rm M_*}$ together. However, the $\Sigma_{\rm SFR}$ (UV and IR values) and $\Sigma_{\rm M_*}$ are plotted in relative units (i.e., the y-axis zeropoint is offset between $\Sigma_{\rm SFR}$ and $\Sigma_{\rm M_*}$).  However, the SFR and stellar-mass surface densities span the same range in logarithm space so the \textit{slope} of the relation between surface density and stellar mass can be directly compared.  
Previous studies have found that the slopes of the correlation between $\Sigma_{\rm eff, M_*}$ and $\Sigma_{\rm 1kpc, M_*}$ and stellar mass are not redshift dependent at $0.5<z<3$, but the intercepts of these correlations change as redshift (e.g., \citealp{Barro2017a}). Thus, we plot low- and high-redshift subsamples separately.  

We see clear correlations between the {stellar mass and the surface density of stellar mass, and between the stellar mass and the surface density of the IR-based SFRs.  However, we see that the} surface density of UV-based SFRs remain relatively constant as stellar mass. 
This is true both for $\Sigma_{\rm eff}$ and $\Sigma_{\rm 1kpc}$. {A reminder that UV-based SFRs is not corrected for dust attenuation (see Section~\ref{sec:Sigma}). }
{To constrain these relations, we fit a linear relation between each surface density and stellar mass using a Gaussian Mixture Model (linmix, \citealp{Kelly2007}). The best-fitted parameters are the median of 400 fitted parameters of random draws from the posterior, and the associated errors are one-half of the difference between the 16th and 84th percentiles of the fitted parameters. This best-fitted lines are shown in Fig. \ref{fig:sigma}, along with 400 random draws from the posterior. }
The slopes of these fitted lines and associated errors are listed in Tab.~\ref{tab:slopes}. 

The $\Sigma_{\rm M_*}$ -- stellar-mass relations derived from literature are indicated by black arrows in Figure~\ref{fig:sigma}.  
{It worth noting that the stellar mass and stellar-mass surface density are not independent. } 
In the top panel of Figure~\ref{fig:sigma}, the black arrow shows the mean size -- stellar-mass relation, $\rm \Delta log(r_e) = 0.22 \Delta log(M_*)$, from \citealp{vanderWel2014}, which gives the surface density of stellar mass within the effective radius growth as $\rm \Delta log(\Sigma_{M_*}) = 0.56 \Delta log(M_*)$. This is consistent with $\rm \Delta log(\Sigma_{M_*}) = 0.60 \pm 0.05 \Delta log(M_*)$ from \citet{Barro2017a}. 
Our best-fit slope of $\Sigma_{\rm eff, M_*}$ {at $z < 1.2$ is} slightly larger than these relations from literature but {only at $\sim$1$\sigma$ significant}. {However, the best-fit slope of $\Sigma_{\rm eff, M_*}$ at $z>1.2$ is significantly larger than these relations. These is similar to the mass-size comparison, where the median F160W size of the high-mass bin for high-redshift subsample is offset from the size-mass relations from \citealp{vanderWel2014}. }
In the bottom panel, the black arrow shows the $\rm \Delta log(\Sigma_{1kpc}) = 0.9 \Delta log(M_*)$ from \citet{Barro2017a}. 
Our $\rm \Sigma_{*, 1kpc}$-$\rm M_*$ relation {at $z < 1.2$ is} nearly identical to that found by \citet{Barro2017a}, {and they are different for the high-redshift subsample}. 

It is interesting that, in both $\Sigma_{\rm eff}$ and $\Sigma_{\rm 1kpc}$ panels, the $\Sigma_{\rm SFR_{IR}}$ increases much faster than the $\Sigma_{\rm SFR_{UV}}$ for both the low- and high-redshift subsamples.  
The difference between $\Sigma_{\rm SFR_{IR}}$ and $\Sigma_{\rm SFR_{UV}}$ suggests that galaxies with higher stellar masses have a higher obscured fraction of star formation in their inner regions. We further explored the obscured fraction of star formation within 1 kpc as function of stellar mass and compare to that integrated over the entire galaxies in Section~\ref{sec:fobs}. 

In addition, the $\Sigma_{\rm SFR_{IR}}$ increases with mass {slightly} faster than the $\Sigma_{M_*}$ for the low-redshift subsample. 
The steeper slope of $\Sigma_{\rm SFR_{IR}}$ is not likely due to differences in morphology because the $R_{\rm eff}$s of the PAH-band are only slightly smaller than those in the F160W in massive galaxies. 
Moreover, the $f_{\rm 1kpc}$ of PAH-band and F160W are largely consistent across the redshift and stellar mass ranges. 
Therefore, this steeper slope is most likely due to the increasing of $\rm SFR_{IR}$ at higher mass.
This result indicates that star-forming massive galaxies, at least at $z \leq 1.2$, experience a faster buildup of their dust content in their inner regions compared to their stellar mass. 
This result could be explained by the fact that more massive galaxies might be intrinsically more capable of dissipative gas accretion toward the center (e.g., \citealp{Dekel2009, Dekel2014}). 
However, the slopes of $\Sigma_{\rm SFR_{IR}}$ and $\Sigma_{M_*}$ are similar for the high-redsfhit subsample. We suspect this is due to the lack of galaxies with stellar mass lower than $10^{9.5} M_\odot$, which seems drive the $\Sigma_{\rm SFR_{IR}}$-mass correlation for the low-redshift subsample.

\subsection{On the Fraction of Obscured Star Formation} \label{sec:fobs}

\begin{figure*}
    \centering
    \includegraphics[width=\textwidth]{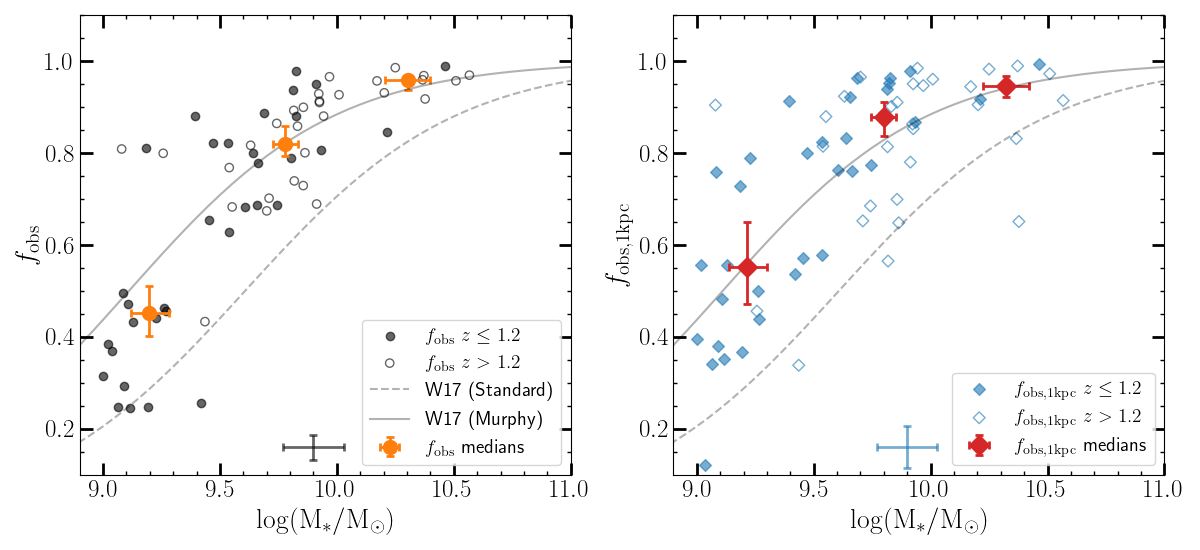}
    \caption{Obscured fraction of star formation as a function of stellar mass integrated over the entire galaxy ($f_{\rm obs}$, \textit{left}) and within 1 kpc ($f_{\rm obs, 1kpc}$, \textit{right}). Solid and open markers are used for galaxies at $z\leq1.2$ and at $z>1.2$, respectively. {The larger orange and red data points with error bars show the median $f_{\rm obs}$ and median $f_{\rm obs, 1kpc}$ versus median stellar mass and associated uncertainties}. 
    {The median of errors on $f_{\rm obs}$, $f_{\rm obs, 1kpc}$ and stellar mass are indicated by black and blue errorbars in the bottom of both panels. }  The $f_{\rm obs}$ {(measured on the whole galaxy)} -- mass relations for galaxies at $z\sim$ 0 -- 2.5 from \citet{Whitaker2017} are shown as the grey lines {in both panels, though it is only a valid comparison to the $f_{\rm obs}$ in the left panel. } }
    \label{fig:fobs}
\end{figure*}

Having estimated both the SFR derived from IR and UV, and SFR surface densities, we can further compare the obscured fraction of star formation integrated over the entire galaxy ($f_{\rm obs}$) and measured within 1 kpc ($f_{\rm obs, 1kpc}$), defined as
\begin{equation}
    f_{\rm obs} = \rm SFR_{IR}/(SFR_{IR} + SFR_{UV}), 
\end{equation}
\begin{equation}
    f_{\rm obs, 1kpc} = \rm \Sigma_{\rm 1kpc,  SFR_{IR}}/(\Sigma_{\rm 1kpc,  SFR_{IR}} + \Sigma_{\rm 1kpc,  SFR_{UV}}), 
\end{equation}
where {$\rm SFR_{IR}$} and $\rm SFR_{UV}$ are the IR-based and UV-based SFR, $\rm \Sigma_{\rm 1kpc, SFR_{IR}}$ and $\Sigma_{\rm 1kpc,  SFR_{UV}}$ are the surface density of IR-based and UV-based SFR within 1~kpc (see Section~\ref{sec:Sigma}). 
In Fig.~\ref{fig:fobs}, we plot the $f_{\rm obs}$ and $f_{\rm obs, 1kpc}$ versus stellar mass {of individual galaxies and the median values in three equally spaced stellar mass bins. The medians and associated errors are obtained from the bootstrap method. }

{Meanwhile, we show t}he $f_{\rm obs}$ -- mass relations for galaxies at $z\sim$ 0 -- 2.5 from \citet{Whitaker2017} {in grey lines as a reference to the relation between the obscured fraction integrated over the entire galaxy and stellar mass. } \citet{Whitaker2017} used Spitzer MIPS 24$\mu$m observations of a mass complete sample at $\rm log(M_*/M_\odot) \geq 9$ with two methods of SFR conversions (``standard'' and ``Murphy''). 
For the ``standard'' method, $\rm SFR_{IR}$ is derived by converting 24 $\mu$m flux densities to total IR luminosity based on \citet{Dale2002} IR SED templates and adopt IR luminosity to $\rm SFR_{IR}$ following \cite{Kennicutt1998}, and $\rm SFR_{UV}$ is derived from rest-frame UV luminosity at 2800\AA\ based on
\citet{Bell2005}. 
For the ``Murphy'' method, $\rm SFR_{IR}$ and $\rm SFR_{UV}$ are derived from 24 $\mu$m flux densities and rest-frame UV luminosity at 1500\AA, respectively, following \citet{Murphy2011}. 
{Our $f_{\rm obs}$ based on the MIRI and rest-frame FUV data favor the observed fractions in the standard conversion from} \citet{Whitaker2017}.

We observe a consistent strong mass-dependence of the obscured fraction of star formation within 1 kpc and on galaxy scale. The former relation has also been shown in Fig.~\ref{fig:sigma}. 
At lower stellar masses $M_* \lesssim 10^{9.5} M_\odot$, the obscured fractions are lower ({$\sim$0.45} for $f_{\rm obs}$ and {$\sim$0.55} for $f_{\rm obs, 1kpc}$) than those in more massive galaxies ($f_{\rm obs}$ and $f_{\rm obs, 1kpc}$ {$\gtrsim0.8$}). 
Therefore, the NUV-band and F160W morphologies of low-mass galaxies should be less affected by the dust attenuation, compared to these morphologies of more massive galaxies. 
This result further supports the assertion that the dust attenuation is not a primary cause for the more extended H$\alpha$ sizes with stellar mass $\rm M_* \sim 10^{9.5} M_\odot$ at $0.5 \leq z \leq 1.7$ \citep[e.g.,][]{Matharu2022}.  Rather this supports ``inside-out'' growth for galaxies in this mass and redshift range. 

Furthermore, we notice that the median $f_{\rm obs, 1kpc}$ of low-mass galaxies ($\rm M_*\leq10^{9.5} M_\odot$) is higher than the median $f_{\rm obs}$ by 10\%, {tentatively} suggesting a higher obscured fraction of star formation within 1 kpc in these low-mass galaxies. This could be {explained by} the more extended profile of NUV-band in low-mass galaxies at $z \leq 1.2$, which lower the surface density of $\rm SFR_{UV, 1kpc}${, and thus, increased the $f_{\rm obs, 1kpc}$}.

\section{Discussion} \label{sec:discussion}

The \jwst/MIRI data have enabled -- for the first time -- the ability to study the mid-IR morphologies of distant galaxies.  This puts these studies on a similar footing as studies of the rest-frame UV and optical morphologies from \hst.   
The comparisons of the  morphological parameters derived from the \jwst/MIRI data and \hst/ACS and WFC3 data (effective radius, S\'{e}rsic index, fraction of light contained within 1 kpc) show differences in the physical distribution of stars (traced by the stellar continuum) and star-formation (traced by the IR--based and UV-based star-formation).  These differences furthermore depend on stellar mass and redshift. 
In particular, our analysis of the surface density of the IR-based SFR, UV-based SFR (and the ratio of the surface density of IR-based and UV-based SFR) and stellar mass reveal that galaxies with higher stellar masses have particularly higher obscured SFR. 

In this section, we first test the robustness of the morphological measurements by considering the effect of the angular resolution of the data (Section~\ref{sec:dis_robust}) and discuss caveats associated with our results (Section~\ref{sec:dis_caveat}). 
We then discuss the implications that our results on the spatially resolved star formation and on galaxy formation, compared to what was known from studies based on rest-frame optical/UV data only (Section~\ref{sec:dis_SF}).

\subsection{The robustness of the morphology measurements} \label{sec:dis_robust}

There are known systematics that can impact the morphological measurements derived from fitting models to the galaxy surface brightness. 
The angular resolution of the image can limit our our ability to measure accurate sizes and S\'ersic indexes (e.g., \citealp{Haussler2007}). 
For galaxies with very compact morphologies (i.e., where the  $R_\mathrm{eff}$ is small compared to the PSF FWHM), the uncertainties on the size and S\'ersic indexes are less robust and more degenerate.  
For galaxies in our sample this will have the most impact on the PAH-band morphologies as the \jwst/MIRI images have a larger PSF FWHM compared to the \hst/ACS and WFC3 data. 
However, we argue this is not a significant factor on our results. The MIRI PSF has a $\sigma$ = FWHM/2.35 that ranges from 0.1--0.3 from 7.7 to 21~\micron.  Figure~\ref{fig:size_n} shows that the majority of the PAH-band effective sizes are larger than 2 kpc, which corresponds to 0.2--0.3\arcsec\ over the redshift range of our sample. This is comparable to -- or larger than -- the PSF and implies that for the majority of our the sample the morpohologies are reasonable resolved (and the \galfit\ results relatively robust). 

\begin{figure}
    \centering
    \includegraphics[width=\columnwidth]{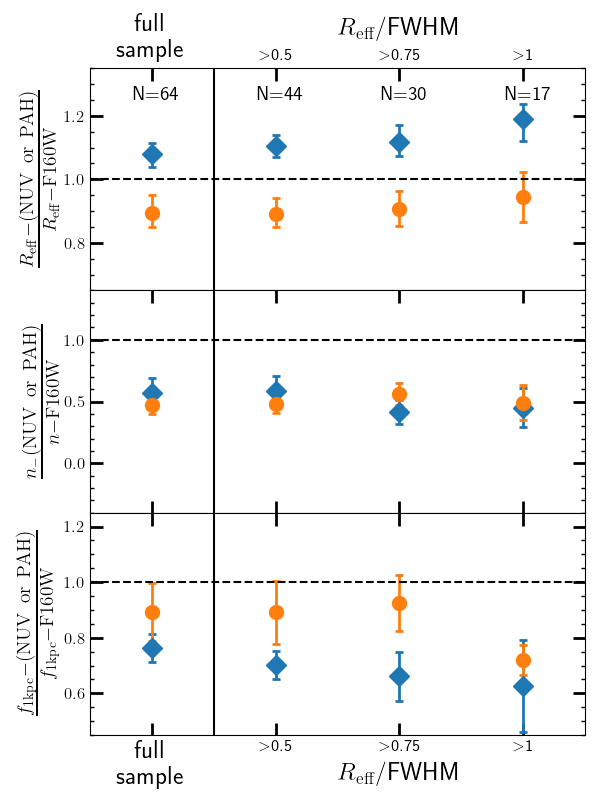}
    \caption{Tests of the effect of image resolution and AGN on the morphological parameter measurements. The \textit{left} panel shows the ratio of the sizes, S\'ersic indexes, and fraction of light contained within 1 kpc between the NUV- and PAH-bands against the  F160W band for the full samples.  The \textit{right} panel shows the results when the sample is divided into galaxies where their sizes are greater than 0.5$\times$, 0.75$\times$, and $1.0\times$ the PSF FWHM of all three bands (PAH-, NUV-bands and F160W). No differences are shown between the full sample and these subsamples, thus, the results on morphological differences do not depend strongly on size. }
    \label{fig:robust}
\end{figure}

Nevertheless, we also test the robustness of the morphology measurements by selecting galaxies with measurements of the $R_{\rm eff}$ that are significantly larger than the PSF. 
For a Gaussian distribution, the $R_{\rm eff}$ of the PSF is 50\% of its FWHM. 
Because morphologies are fitted with different S\'{e}rsic indexes, we then subdivide these galaxies into samples with $R_{\rm eff}$/FWHM $>$ 0.5, 0.75, and 1 in all three of the NUV-, F160W-, and PAH-bands. 
The median $R_{\rm eff}$, $n$ and $f_{\rm 1kpc}$ ratios of PAH- and NUV-bands to F160W for these subsamples are shown in Fig.~\ref{fig:robust}. 
The average differences between these morphologies of PAH-/NUV-band and F160W of the full sample and these subsamples remain unchanged. As we found above for the full sample, the PAH-band sizes are slightly smaller, with lower $n$, and similar $f_\mathrm{1~kpc}$ as F160W, and the NUV-band sizes are slightly larger, with lower $n$, and lower $f_\mathrm{1~kpc}$ compared to the values in the F160W band. 
Quantitatively, the $R_{\rm eff}$ and $f_{\rm 1kpc}$ of PAH-band/F160W for galaxies with $R_{\rm eff}$/FWHM$>$1 shift to somewhat larger and more extended region.
This is mostly likely due to a selection effect. We are preferentially {selecting} galaxies with larger PAH, because this PSF selection dominantly act on the PAH, given the relatively large PSF of MIRI image. 
We therefore conclude that the trends in the data are not driven by differences in the angular resolution of the different bands.

\subsection{Caveats} \label{sec:dis_caveat}

{In this section, we discuss some caveats that might affect our results. Firstly, our results on the morphology---stellar-mass relations are based on a single redshift/mass binning that does not account for uncertainties on stellar mass, redshift and sample incompleteness. Our binning method is chosen to have a similar number of galaxies in each bin in order to increase the statistical power in each bin.}

{We have tested the effect of using different binning methods, but we find these do not seriously impact our results.  For the low-redshift subsample, if we bin galaxies into three bins of stellar mass, $\rm log(M_*/M_\odot)$ $<$ 9.35, $\sim$9.35--9.70, and $>$9.70), then the median sizes of PAH-band remain $\sim$10\% smaller than those of F160W for galaxies in $\rm log(M_*/M_\odot)\sim$ 9.35--9.70 and $>$ 9.70 at 3.5$\sigma$ and 2$\sigma$ significant levels. Meanwhile, the median size of NUV-band remain $\sim$10\% larger than that of F160W for galaxies in $\rm log(M_*/M_\odot)\sim$ 9--9.35 at a 2$\sigma$ significant level. For the high-redshift subsample, if we bin galaxies into three bins in stellar mass, $\rm log(M_*/M_\odot)$ $\sim$ 9.50--9.75, $\sim$9.75--10, and $>$ 10, then we see larger errors on the median size ratios of PAH-band and NUV-band to F160W, similar to Fig.~\ref{fig:size_mass_z}. Thus, using slightly different binning methods does not affect our main results significantly. } 

{Regarding the sample incompleteness, we lack galaxies with high stellar mass at low redshift, mostly due to the small survey area.  We used only two MIRI pointings that have UV coverage with UVCANDELS), which have a total area of $\lesssim$ 5 arcmin$^2$. We also are lacking galaxies with low stellar mass at high redshift, because they appear fainter in brightness and are excluded due to \galfit\ failures.  This restricts our sample to the stellar-mass range studied above. In addition, as mentioned in Section~\ref{sec:sample}, our sample selection might bring a potential bias toward SFGs with moderate dust and against very dusty SFGs because of excluding NUV-undetected sources.  Both of these potential biases will be rectified from larger-area, and deeper imaging with future MIRI observations.}

{The second caveat is related to our estimate of the IR SFR. We use the mid-IR to anchor the total IR, which could lead to uncertainties in the total IR luminosity and IR SFR. }
{We derive the total IR luminosities using CIGALE to model the full UV, optical, and IR SED and with fluxes span from UV to Mid-IR (as presented in Section~\ref{sec:cigale}). These SED fitting include limits from \spitzer/MIPS (or in a few cases detections), which restricts the upper limit of IR luminosities. Nevertheless, the lack of longer wavelength data could lead to additional systematic uncertainties in our IR luminosities \citep{Liu2018, Jin2018}. This could be improved by stacking the MIRI-detected galaxies in the MIPS data, and also in \herschel/PACS and SPIRE data that exist in these fields, which would improve the average relation between the mid-IR flux densities and the full shape of the far-IR SED. Such analysis is beyond the scope of this work, but we expect to pursue it in a future study.  }

\subsection{Implications for Spatially Resolved Star-Formation in Distant Galaxies} \label{sec:dis_SF}

In this paper we have measured the morphologies of MIRI-detected SFGs in the rest-frame NUV, PAH emission, and stellar continuum.  This allows us to account for where stars are forming as traced by the NUV and PAH emission, compared to where they have formed, as traced by the stellar continuum.  Here we interpret these results in terms of how galaxies grow with time.  

First, it is interesting that the S\'{e}rsic indexes derived from the PAH emission and NUV-band data favor exponential disks, or shallower profiles, with $n \lesssim 1$.  
In contrast the S\'ersic indexes of the stellar continuum traced by the F160W data favor slightly larger values, with $n \gtrsim 1$. 
One interpretation from this is that star-formation proceeds in a disk-like fashion. The stellar continuum, which is the integral of the history of star-formation, shows that star-formation must have first occurred in the central regions, with higher densities ($\Sigma_\ast$), and subsequently quenched such that the profile of the stellar mass remains more concentrated than the ongoing star-formation.  

Second, we consider the size evolution, where our analysis of the NUV-band morphologies suggests that unobscured star-formation sizes are larger and more extended than the stellar continuum. 
At first glance, this is consistent with results measured from H$\alpha$ at $z\sim0.5-3$ \citep{Nelson2016b, Wilman2020, Matharu2022}. 
These earlier studies found that the sizes of the H$\alpha$ emission are exceed that of the stellar continuum suggesting ongoing star-formation activity at preferentially larger radii than existing stars (i.e., galaxies are forming ``inside--out''). Although, these studies may be impacted by effects of dust attenuation \citep[e.g.,][]{Nelson2016a}.

It is instructive to compare the results on the H$\alpha$ and NUV sizes of galaxies.
\citet{Matharu2022} found that the mean H$\alpha$ effective radius is $1.2\pm0.1$ times larger than that of the stellar continuum for SFGs with $9.25 < \rm log(M_*/M_\odot) < 9.70$ in the redshift range $0.5\lesssim z\lesssim 1.7$ (similar to that of our sample here). No difference is seen between the H$\alpha$ and stellar continuum sizes in less massive galaxies ($8.96<\rm log(M_*/M_\odot) < 9.25$), but the H$\alpha$ sizes are slightly larger than the stellar continuum for more massive galaxies ($9.70<\rm log(M_*/M_\odot) < 11.3$). 
We similarly observe that the median NUV-band effective radius is {$1.15_{-0.04}^{+0.09}$} times larger than that of the stellar continuum in galaxies with $9 < \rm log(M_*/M_\odot) < 9.45$ and at $z\leq 1.2$.  Therefore the NUV and {H$\alpha$} morphologies provide reinforcing evidence for this ``inside-out'' growth, at least for galaxies at lower stellar masses.  

On the other hand, we find that the PAH emission on average is slightly \textit{smaller} ($\sim$10\%) than the stellar continuum in galaxies with $\rm M_* \gtrsim 10^{9.5}M_\odot$ at $z \leq 1.2${, and the sizes of PAH emission and stellar continuum are similar in  galaxies with lower stellar masses. } (see Fig~\ref{fig:size_mass_z}). 
For galaxies at higher stellar masses, the fact that the sizes of the PAH emission are smaller than the stellar continuum implies there is an increase in dust-obscured star formation in the central regions of galaxies, and that this increases with stellar mass. This result is consistent with the stellar mass dependence on the extent of dust attenuation using Balmer decrements of H$\alpha$/H$\beta$ \citep{Nelson2016a} and UV continuum \citet{Tacchella2018}. 

{Previous studies have explored the evolution of the sizes of the rest-frame far-IR continuum measurements from ALMA as a tracer of the obscured-star formation. These studies generally revealed that the dust-obscured star formation is much more compact compared to the stellar continuum} \citep{Hodge2015, Simpson2015, Chen2017, CalistroRivera2018, Gullberg2019, Hodge2019, Lang2019, Tadaki2017a, Tadaki2017b, Fujimoto2017, Tadaki2020, Cheng2020, Gomez-Guijarro2022}. 
{However, these ALMA results have been limited almost entirely to galaxies more massive than $\rm M_* \gtrsim 10^{10.5} M_\odot$, which is even higher than our highest mass bin. } 
{In our highest mass bin ($\log(\rm M_*/M_\odot) = 10-10.5$ at $z > 1.2$), we find that the sizes of galaxies in the PAH-band are similar to those in the F160W. This is potentially inconsistent with the ALMA results, even we consider a 10\% decrease in the size ratio of PAH-band to F160W between galaxies with $\log(\rm M_*/M_\odot) = 10-10.5$ and galaxies with $\log(\rm M_*/M_\odot) > 10.5$ assumed based on the decrease observed in the low-redshift subsample. While, our highest mass bin contains only 15 galaxies, with a large spread in the $R_\mathrm{eff}$ ratio of PAH-band to F160W (see Figure~\ref{fig:size_mass_z}). } 
{Based on the trend between the galaxy sizes in the PAH-band and F160W from low-redshift subsample, we expect this is evidence that the dust-obscured star-formation becomes more compact with increasing mass, which could connect the MIRI-based sizes here with the ALMA ones in the literature. Although, it might be possible that the physical scales measured from ALMA are affect by the interferometry that missed the diffuse Far-IR emission and underestimate the extent of Far-IR emission. These issues will be discussed further by Magnelli et al., \textit{in prep}. }

Therefore, the interpretation of the NUV and PAH sizes are enigmatic. 
One possibility is that the NUV and PAH emission trace star-formation on different timescales (where the NUV traces the direct continuum from OB-type stars with lifetimes of $\sim 10-100$~Myr while the PAH and IR emission can include heating from longer-lived stellar stars with lifetimes of 500~Myr to 1 Gyr \citep{Salim2009,Salim2020,Kennicutt2012}.  It is therefore plausible that if galaxies grow and quench via an ``inside-out'' process, the PAH appearances would lag behind that of the UV. This can account for the growth of galaxies, at least at lower stellar masses. 

However, at higher stellar masses, our PAH results do not favor this ``inside-out'' scenario as that PAH-based effective radii become smaller than that of the NUV or of the stellar continuum. This favors an interpretation that star-formation is increasing obscured, and more centrally concentrated. This is apparent by the fact that the $\rm SFR_{IR}$ and $\rm \Sigma_{IR}$ is higher than $\rm SFR_{UV}$ and $\rm \Sigma_{UV}$ for galaxies with $log(M_*/M_\odot) \gtrsim 9.5$, and that the fraction of the IR SFR increases to higher stellar mass (Figure~\ref{fig:fobs}). 
{The observation that the star-formation in SFGs becomes more compact as the galaxies grow in mass has important ramifications for their evolution. For example, it will be important to test if the SFR surface densities versus galaxy stellar mass connect to predictions for bulge growth in galaxies (including if galaxies ``compactify'', e.g., \citealp{Dekel2014, Ceverino2010, Zolotov2015, Tacchella2016a}), and if galaxy quenching occurs when stellar mass or the stellar mass surface density reach some critical value. This can be tested by future studies with larger samples, particularly for larger samples of more massive galaxies with MIRI coverage.  }

Another possible explanation might be that the contribution by older stellar populations (as opposed to the forming ones) to the excitation of the PAH grains is proportionally larger in more massive galaxies. Thus, in their central regions, a larger fraction of the PAH emission are being excited by existing stars in massive galaxies than that in lower mass galaxies, which is shown as higher $\rm \Sigma_{IR}$. 
Other reasons also include that the complex morphology of massive high-redshift galaxies makes it difficult to reproduce the mass/light distribution with a single-component model, such as compact bulges being formed \citep{Costantin2021, Costantin2022a, Guo2022}.

Nevertheless, from various simulations, the effect of dust obscuration has been pointed out as one of the reasons for the discrepancy in the intrinsic and observed size of massive galaxies observed in rest-frame optical wavelength \citep{Costantin2022b} and also at rest-frame Far-UV \citep{Marshall2022, Roper2022}. 
This is supported by previous studies which found the half-mass radii is, on average, smaller than the half-light radii for galaxies with stellar mass $\gtrsim10^{10} M_\odot$ at $1.0 \leq z \leq 2.5$ \citep{Suess2019}. 
The PAH-band size -- mass relation from this paper seems to follow the intrinsic size -- mass relation from these simulations that decreases as increasing stellar mass. 
Thus, the PAH or MIR morphology could help in correcting the rest-frame optical/FIR and rest-frame UV or H$\alpha$ morphology and to obtain an intrinsic size -- mass relation, which, in turn, can inform the formation and quenching mechanisms.

\section{Summary}\label{sec:summary}

We use imaging from \jwst/MIRI covering 7.7--21~\micron\ to study the morphologies of galaxies based on their PAH emission.  The MIRI data resolve angular structures in galaxies on scales of 0.1-0.3\arcsec\ and allow for the first time a quantitative measurement of galaxy morphologies in the mid-IR for the first time.  

We compare the morphologies of {64} star-forming galaxies at $0.2 < z < 2.5$ with stellar mass $\rm log(M_*) > 10^9\ M_\odot$ in their in PAH-band, NUV-band and F160W using data form \jwst\ MIRI, \hst\ ACS/F435W or ACS/F606W and \hst\ WFC3/160W. 
These three bands trace the profiles of obscured-, unobscured-star formation, and stellar continuum, respectively. 
We found there are differences in the galaxy morphologies in their PAH-, NUV-bands and F160W, which depend on stellar mass and redshift.  Our results are summarized as the following.  

\begin{itemize}
    
    \item Comparing to sizes of galaxies in the stellar continuum traced by F160W to the obscured-star formation traced by the PAH-band, we find that the latter are slightly smaller size (in $R_\mathrm{eff}$) with a similar fraction of light within 1 kpc (similar in $f_\mathrm{1kpc}$) for galaxies with $\rm log(M_*) \gtrsim 10^{9.5} M_\odot$ at $z\leq1.2$. 
        
    \item Comparing the sizes of galaxies in the stellar continuum traced by F160W to the unobscured-star formation traced by the NUV-band, we find that the latter is larger (in $R_\mathrm{eff}$) and more extended (lower in $f_\mathrm{1kpc}$) for galaxies at $z\leq1.2$, but the NUV-band and F160W values are more similar for galaxies at $z\leq1.2$.    
    
    \item The average S\'{e}rsic indices of galaxies measured in their NUV- and PAH-bands are similar with ($\langle n\rangle \sim 0.7$).  These are both smaller than the average S\'ersic indices measured in the F160W band ({$\langle n\rangle =1.1$}). These results reveal that the stellar continuum prefer a disk-like profile, while the obscured- and unobscured-star foramtion follow a flatter profile (within the effective radius). 
    
    \item  We estimate the surface density of the SFR and stellar mass density, by combining information from the morphological tracers of star formation (the NUV- and PAH-band) with the total SFR (from UV and far-IR data), and using morphological tracers of the stellar continuum (from the F160W data) with the total stellar mass.    We find that the surface density of the SFR derived from IR increases faster with increasing stellar mass (i.e., it has a steeper slope) than the surface density of the SFR derived from UV. We interpret this as evidence that galaxies with higher stellar masses have preferentially higher amounts of centrally concentrated dust-obscured star-formation. 
    
    \item The surface density of the SFR derived from the IR also increases with stellar mass with a steeper slope than the surface density of stellar mass at $z\leq1.2$, suggesting that massive SFGs, at least at $z \leq 1.2$, experience a faster buildup of their dust content in their inner regions compared to their stellar mass. At $z > 1.2$, the relation between the IR-based SFR surface density and mass has a similar slope as the stellar mass surface density and mass, possible due to the lack of low-mass galaxies with $\rm M_* \lesssim 10^{9.5} M_\odot$. 
    
    \item We study the fraction of obscured star formation (defined as the ratio of the IR-based SFR to the total IR + UV SFR), both measured within 1 kpc and integrated over the entire galaxy.   We find a strong correlation between the fraction of obscured star formation and stellar mass, both for the fraction measured within 1 kpc, and the fraction integrated over the entire galaxy. The obscured fraction is generally lower for lower mass galaxies ({$f_\mathrm{obs}\sim$0.5} for masses of $10^{9-9.5}~M_\odot$) than for more massive galaxies ($f_\mathrm{obs} \gtrsim0.7$ for masses of {$\gtrsim 10^{9.5}~M_\odot$}), suggesting the dust attenuation is not a primary cause for the more extended NUV in these low-mass galaxies. 
    
\end{itemize}

This paper demonstrates the capability of MIRI data to reveal the morphology of dust-obscured star forming regions in star-forming galaxies down to $\rm M_*\sim10^{9}M_\odot$. 
We expect future MIRI surveys will provide a more complete sample to study the morphology of obscured star formation and dust attenuation, which can inform the spatially-resolved growth of galaxies and the intrinsic size -- mass relation. 

\begin{acknowledgements}
We acknowledge the hard work of our colleagues in the CEERS and UVCANDELS collaboration and everyone involved in the \jwst\ mission.   
This work benefited from support from the George P. and Cynthia Woods Mitchell Institute for Fundamental
Physics and Astronomy at Texas A\&M University. 
CP thanks Marsha and Ralph Schilling for generous support of this research. 
RAW acknowledges support from NASA JWST Interdisciplinary Scientist grants NAG5-12460, NNX14AN10G and 80NSSC18K0200 from GSFC.  This work
acknowledges support from the NASA/ESA/CSA James Webb Space Telescope
through the Space Telescope Science Institute, which is operated by
the Association of Universities for Research in Astronomy,
Incorporated, under NASA contract NAS5-03127. Support for program
No. JWST-ERS01345 was provided through a grant from the STScI under
NASA contract NAS5-03127.
{This work is based on observations with the NASA/ESAHubble Space Telescope obtained at the Space Telescope Science Institute, which is operated by the Association of Universities for Research in Astronomy, Incorporated, under NASA contract NAS5- 26555. Support for Program number HST-GO-15647 was provided through a grant from the STScI under NASA contract NAS5-26555. }
\end{acknowledgements}

\bibliography{ref}{}
%\bibliographystyle{aasjournal}



\end{document}